\documentclass[a4paper,fleqn,usenatbib]{mnras}

\usepackage[T1]{fontenc}
\usepackage{ae,aecompl}

\usepackage{graphicx}	
\usepackage{amssymb}	
\usepackage[greek,english]{babel}
\usepackage{amssymb}
\usepackage{amsmath}
\usepackage{times} 
\usepackage{mathptmx}
\usepackage{listings}
\usepackage{gensymb}
\usepackage{amsmath,xspace}
\usepackage{pdflscape}
\let\oldAA\AA
\renewcommand{\AA}{\text{\normalfont\oldAA}}
\newcommand{\rmctwentyseven}{RMC$\,$127\xspace}
\newcommand{\rmcfortythree}{RMC$\,$143\xspace}

\newcommand{\qso}{QSO$\,$J0635-7516\xspace}
\newcommand{\bpcal}{QSO$\,$J0538-4405\xspace}
\newcommand{\fluxcal}{Pictor A\xspace}

\DeclareSymbolFont{newfont}{OML}{cmm}{m}{it}
\DeclareMathSymbol{\Varrho}{3}{newfont}{37}

\usepackage{color}

\usepackage{hyperref}
\hypersetup{
  bookmarks=false,         
  bookmarksopen=false,     
  pdffitwindow=true,       
  pdftitle={New ATCA, ALMA and VISIR observations of the candidate LBV SK-67266 (S61): the nebular mass from modelling 3D density distributions},    
  pdfauthor={Agliozzo et al. 2016},   
  colorlinks=true,         
  linkcolor=blue,          
  citecolor=blue,          
  urlcolor=blue,           
  linktocpage=true
}

\def\Ha{H{$\alpha$}}
\newcommand{\Msun}  {\ensuremath{M_\odot}} 

\newcommand\rhocube {\textsc{Rhocube}}   
\newcommand\xoff {\ensuremath{\texttt{xoff}}}
\newcommand\yoff {\ensuremath{\texttt{yoff}}}
\newcommand\dif  {\hbox{${\rm d}$}}

\newcommand\rhoxyz {\ensuremath{\Varrho\!\!(x,y,z)}}
\newcommand\xyz {\ensuremath{(x,y,z)}}

\title[ATCA, ALMA and VISIR observations of S61]{New ATCA, ALMA and VISIR observations of the candidate LBV SK$\,$-67$\,$266 (S61): the nebular mass from modelling 3D density distributions}
\author[C. Agliozzo]{C. Agliozzo,$^{1,2}$\thanks{E-mail: c.agliozzo@gmail.com}
R. Nikutta,$^{3,4}$ 
G. Pignata,$^{2,1}$
N. M. Phillips,$^{5,6}$
A. Ingallinera,$^{7}$
C. Buemi,$^{7}$
\newauthor
G. Umana,$^{7}$
P. Leto,$^{7}$
C. Trigilio,$^{7}$
A. Noriega-Crespo,$^{8}$
R. Paladini,$^{9}$
F. Bufano,$^{7}$
\newauthor
F. Cavallaro$^{7,10,11}$
\\
$^{1}$Millennium Institute of Astrophysics (MAS), Nuncio Monse{\~{n}}or S{\'{o}}tero Sanz 100, Providencia, Santiago, Chile\\
$^{2}$Departamento de Ciencias Fisicas, Universidad Andres Bello,  
Avda. Republica 252, Santiago, 8320000, Chile\\
$^{3}$National Optical Astronomy Observatory, 950 N Cherry Ave, Tucson, AZ 85719, USA\\
$^{4}$Instituto de Astrof{\'{\i}}sica, Facultad de F{\'{i}}sica, Pontificia Universidad Cat{\'{o}}lica de Chile, Casilla 306, Santiago 22, Chile\\
$^{5}$European Southern Observatory, Alonso de C\'{o}rdova 3107, Vitacura, Santiago, Chile\\
$^{6}$Joint ALMA Observatory, Alonso de C\'{o}rdova 3107, Vitacura, Santiago, Chile\\
$^{7}$INAF-Osservatorio Astrofisico di Catania, Via S. Sofia 78, 95123 Catania, Italy\\
$^{8}$Space Telescope Science Institute 3700 San Martin Dr., Baltimore, MD, 21218, USA\\
$^{9}$Infrared Processing Analysis Center, California Institute of Technology, 770 South Wilson Ave., Pasadena, CA 91125, USA\\
$^{10}$CSIRO Astronomy and Space Science, PO Box 76, Epping, NSW 1710, Australia\\
$^{11}$International Centre for Radio Astronomy Research, Curtin University, Bentley, WA 6102, Australia\\
}

\date{Accepted 2016 November 15. Received 2016 November 14; in original form 2016 October 13}

\pubyear{2016}

\begin{document}
\label{firstpage}
\pagerange{\pageref{firstpage}--\pageref{lastpage}}
\maketitle

\begin{abstract}
  We present new observations of the nebula around the Magellanic candidate 
  Luminous Blue Variable S61. These comprise high-resolution
  data acquired with the Australia Telescope Compact Array (ATCA), the
  Atacama Large Millimetre/Submillimetre Array (ALMA), and VISIR at
  the Very Large Telescope (VLT). The nebula was detected only in the
  radio, up to 17 GHz.  The 17 GHz ATCA map, with 0.8 arcsec
  resolution, allowed a morphological comparison with the
  $\rm H \rm \alpha$ {\it Hubble Space Telescope\/} image. The radio
  nebula resembles a spherical shell, as in the optical. The spectral
  index map indicates that the radio emission is due to free-free
  transitions in the ionised, optically thin gas, but there are hints
  of inhomogeneities. We present our new public code \rhocube\ to
  model 3D density distributions, and determine via Bayesian inference
  the nebula's geometric parameters. We applied the code to model the
  electron density distribution in the S61 nebula. We found that
  different distributions fit the data, but all of them converge to
  the same ionised mass, $\sim0.1\,\rm \Msun$, which is an order of magnitude smaller than previous estimates. We show how the nebula
  models can be used to derive the mass-loss history with
  high-temporal resolution. The nebula was probably formed through stellar winds, rather than eruptions. From the ALMA and VISIR
  non-detections, plus the
  derived extinction map, we deduce that the infrared emission
  observed by space telescopes must arise from extended, diffuse
  dust within the ionised
  region. 
\end{abstract}

\begin{keywords}
stars: individual (SK$\,$-67$\,$266)  -- stars: mass-loss -- stars: circumstellar matter -- stars: massive -- radio continuum: stars -- methods: statistical
\end{keywords}



\section{Introduction}
Luminous Blue Variables (LBVs) are evolved massive stars ($>20\,\rm
\Msun$), intrinsically bright ($L\sim 10^{5}-10^{6.3}\,\rm
L_{\odot}$) and  hot (O, B spectral type). They are unstable and exhibit
spectroscopic and photometric variability. During the LBV variability cycle they can resemble a cooler supergiant of spectral type A or F, and show visual magnitude variations over a wide range of amplitudes and timescales \citep[as discussed and reviewed by][]{1994HD, 2001vanGenderen}. Because of
their instability, they suffer mass-loss at high rate
($\dot{M}\gtrsim10^{-5}\,\rm \Msun\,yr^{-1}$) and form
circumstellar nebulae. The mechanism that causes this instability is
still poorly understood. To explain the common ``S Doradus type'' outbursts (with visual magnitude variations of 1-2 mag on timescales of years), changes of the photospheric physical conditions have been invoked. This variation of the photospheric physical conditions is caused by a change of the wind efficiency due to variation of the ionisation of Fe, which is the main carrier of line-driven stellar winds. This mechanism is known as the ``bi-stability jump'' \citep[explained by][]{1990Pauldrach,1995Lamers}, a predicted effect of which is mass-loss variability \citep{1999Vink}. The observational mass-loss rates estimated from different indicators (e.g. UV and optical emission lines, radio free-free emission) have often been discrepant, most of the time depending on whether clumped or unclumped wind models were assumed. For example, \citet{2006Fullerton} found that mass-loss rates estimated from \mbox{P\,{\sc v}} lines in clumpy stellar winds of O stars are systematically smaller than those obtained from squared electron density diagnostics (e.g. H$_{\alpha}$ and radio free-free emission) with unclumped wind models, resulting in empirical mass-loss rates overestimated by a factor 10 or more. The implication is that  
line-driven stellar winds are
not sufficient to strip off quickly the H envelope, before they
evolve to Wolf-Rayet (WR) stars \citep{1976Conti}. Enhanced mass-loss was therefore proposed to reduce the stellar mass, possibly through short-duration eruptions or explosions \citep{1994HD, 2006S&O}. Subsequently, \citet{2007Oskinova} showed that if macro-clumping (instead of optically thin, micro-clumping) is taken into account, \mbox{P\,{\sc v}} lines become significantly weaker and lead to underestimation of the mass-loss rate. Finally, \citet{2012V&G} showed that for moderate clumping (factor up to 10) and reasonable mass-loss rate reductions (of a factor of 3) the empirical mass-loss rates agree with the observational rates and, more importantly, with the model-independent transition mass-loss rate, which is independent of any clumping effects. The implication of this is that eruptive events are not needed to make WR stars. 

The mechanism that triggered the ``giant eruptions'' (with visual magnitude changes larger than 2 mag) witnessed in the 17th (P Cygni) and in the 19th century ($\eta$ Carinae) in our Galaxy is still unknown, but some scenarios involving hydrodynamic (sub-photospheric) instabilities, rapid rotation and close binarity have been proposed \citep[e.g. ][and ref. therein]{1994HD}. The presence of nebulae in most of the known objects \citep[e.g.,][]{1994HD,2001vanGenderen, Clark2005} suggests that these are a common aspect of the LBV behaviour \citep{2008Weis}. 

Given the short duration of the LBV phase ($10^{4}-10^{5}\,\rm yr$),
combined with the rapid evolution of massive stars, LBVs are rare:
only a few tens of objects in our Galaxy and in the Magellanic Clouds
(MCs) \citep{2012HD} satisfy the variability criteria 
coupled with high mass-loss rates \citep{1994HD}. Nevertheless, based
on the discovery of dusty ring nebulae surrounding luminous stars, the
number of Galactic candidate LBVs (cLBVs) has increased recently to 55
\citep{2010Gvaramadze, 2011Wachter, 2012Naze}. A few tens of confirmed LBVs have been discovered in farther galaxies \citep[e.g. M31, M33, NGC2403][and ref. therein]{2016Humphreys}.

LBV ejecta are the
fingerprints of the mass-loss phenomenon suffered by the star. The LBV
nebulae (LBVNe) observed in our Galaxy usually consist of both gas and
dust. Previous studies of known Galactic LBVs at radio wavelengths,
which trace the ionised component, estimated the masses of the nebulae
and their current mass-loss rates \citep[e.g.][]{2002DuncanWhite,
  2005Lang, 2005Umana, 2010Umana, 2011Umana, 2012Umana, 2010Buemi,
  2012Agliozzo, 2014Agliozzo, 2012Paron, 2016Buemi}.  On the other
hand, IR observations revealed that the dust is often distributed
outside of the ionised region, indicative of mass-loss episodes of  
different epochs and/or that the nebulae are ionisation-bounded
\citep[e.g., G79.29$+$0.46, G26.47$+$0.02, Wray 15-751, AG
Car,][]{2010Kraemer, 2010Esteban, 2011UmanaG79, 2012Umana, 2013Vam,
  2015Vam}. These studies show that multi-wavelength, high spatial
resolution observations are needed to determine the mass-loss history
and the geometry associated with massive stars near the end of their
lives \citep{2011Umana}. This information is fundamental to test
evolutionary models. However, some of the parameters associated with
the mass-loss still have large uncertainties, partly due to imprecise
distance estimates, but also due to arbitrary assumptions about the
nebula geometry. 

To understand the importance of eruptive mass-loss in different metallicity environments, we 
observed at radio-wavelengths a sub-sample of LBVs
in the Large Magellanic Cloud (LMC), that has a lower metal content
($Z <0.5\,\rm Z_{\odot} $) than the Milky Way. We selected this sub-sample based on the presence of an
optical nebula \citep{Weis2003}. In \citet{2012Agliozzo} (hereafter
Paper I) we presented for the first time radio observations, performed with ATCA 
at 5.5 and \hbox{9\,GHz}. We detected the radio
emission associated with LBVs \rmctwentyseven, \rmcfortythree and
candidates LBVs (cLBVs) S61 and S119. 
In this work we present the most recent observations of cLBV
S61, covering a larger spectral domain and including Australia
Telescope Compact Array (ATCA), Atacama Large Millimeter/Submillimeter
Array (ALMA), and Very Large Telescope (VLT) VISIR data. The goals of this work are: \emph{(i)} to introduce a quantifiable and objective method for
determining the nebular mass via Bayesian estimation of geometrical
nebula parameters; \emph{(ii)} to derive the mass-loss history with
high temporal resolution; \emph{(iii)} to compare the nebular properties of S61 with similar Galactic LBVNe, with respect to the nebular mass, kinematical age of the nebula and dust production.

S61 (also named SK$\,$-67$\,$266 and AL$\,$418) is only a candidate
LBV because, since its first observations \citep{1977Walborn}, it has
not shown both spectroscopic and photometric
variability. 
The star was classified as luminous supergiant (Ia) spectral type
O8fpe. 
Originally \rmctwentyseven also belonged to this class, until it
entered a state of outburst \citep[between
1978--1980,][]{1982Walborn}, during which the Of features disappeared
and the spectrum evolved through an intermediate B-type to a peculiar
supergiant A-type. In the meantime, Of-type emission was discovered
during a visual minimum of the LBV AG Car \citep{1986StahlO}, the
Galactic twin of \rmctwentyseven. All these findings suggested that
Ofpe stars and LBVs are physically related \citep[e.g.,][]{1986StahlO,
  1989Bohannan, 1998Smith}, and Ofpe supergiant stars are now
considered quiescent
LBVs. 
In this paper we will focus our attention on S61 for which
\citet{1997Crowther} derived the following stellar parameters:
$T_{eff}=27600\,\rm K$, $\log L/L_\odot=5.76$,
$\dot M=1.1\times10^{-5}\,\rm \Msun\,yr^{-1}$ and
$v_{\infty}=250\ \rm km\,s^{-1}$.  The paper is organised as follows:
in Section~\ref{sec:data} we present the new observations and data
reduction; we describe the nebula around S61, its morphology, flux
densities and spectral index (Section~\ref{sec:analysis}); we present
our new public code \rhocube\ \citep{rhocube} to model 3D density
distributions, and derive via Bayesian inference the geometrical
nebula parameters (Section~\ref{sec:model}). From the marginalized
posteriors of all parameters obtained from fitting the 
\hbox{9\,GHz} and \hbox{17\,GHz} maps of S61, we estimate the
posterior PDF of the ionised mass contained in the nebula. In
Section~\ref{sec:discuss} we also show a method to derive the
mass-loss history with high temporal resolution and we compare it with S61's empirical mass-loss rate. We discuss the
derived extinction maps and interpret them together with the mid-IR
and ALMA non-detections. Finally, in Section~\ref{sec:summary} we
summarise our
results. 

\section{Observations}
\label{sec:data}

\subsection{ATCA observations and data reduction}
 
\begin{figure}
  \centering
  \begin{minipage}{1\linewidth}
    \includegraphics[trim=.5cm 0cm .2cm 2cm, clip=true,width=\textwidth]{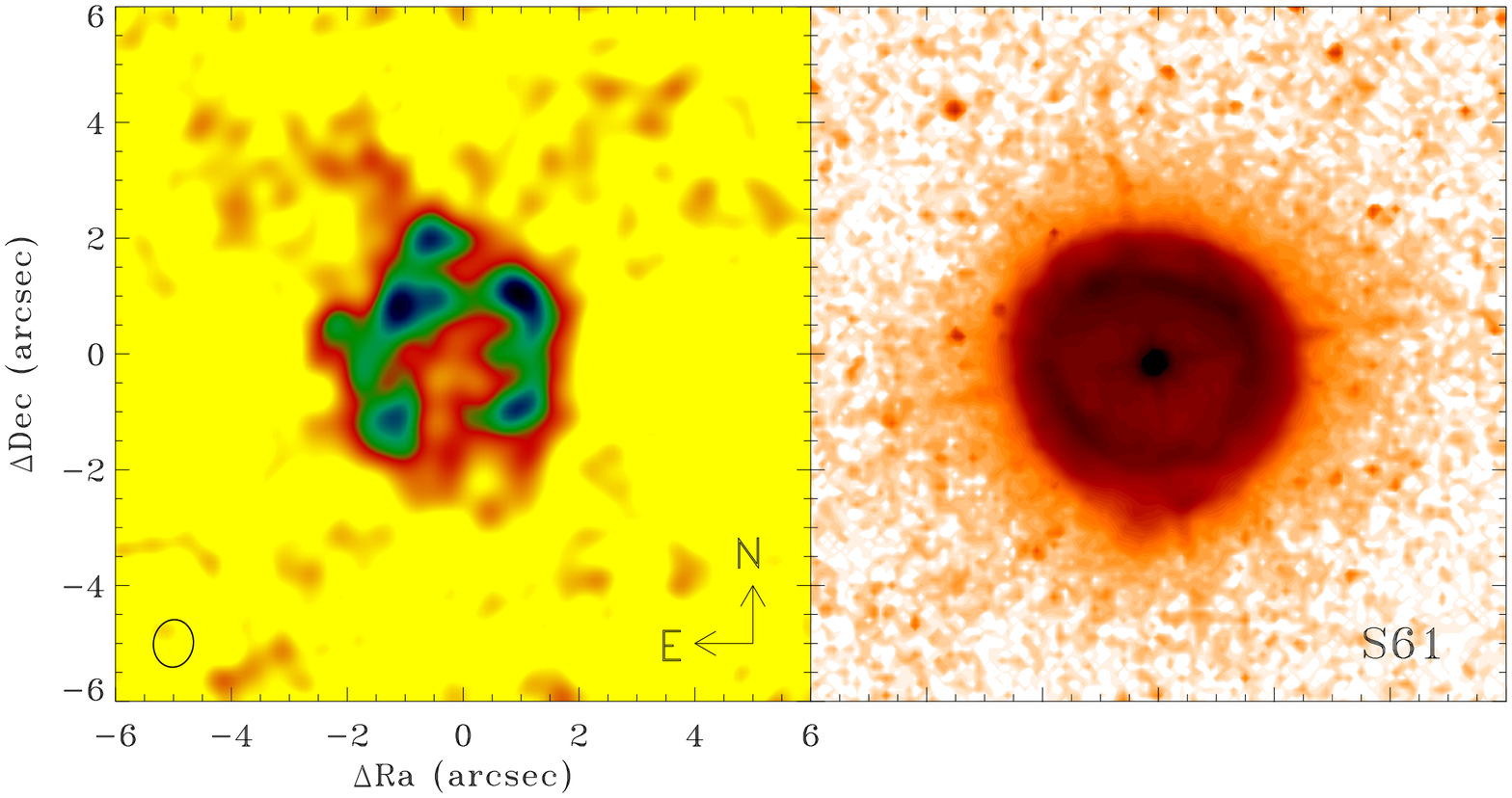}
    \begin{center}\includegraphics[trim=.6cm .1cm .6cm .6cm, clip=true,width=1\textwidth]{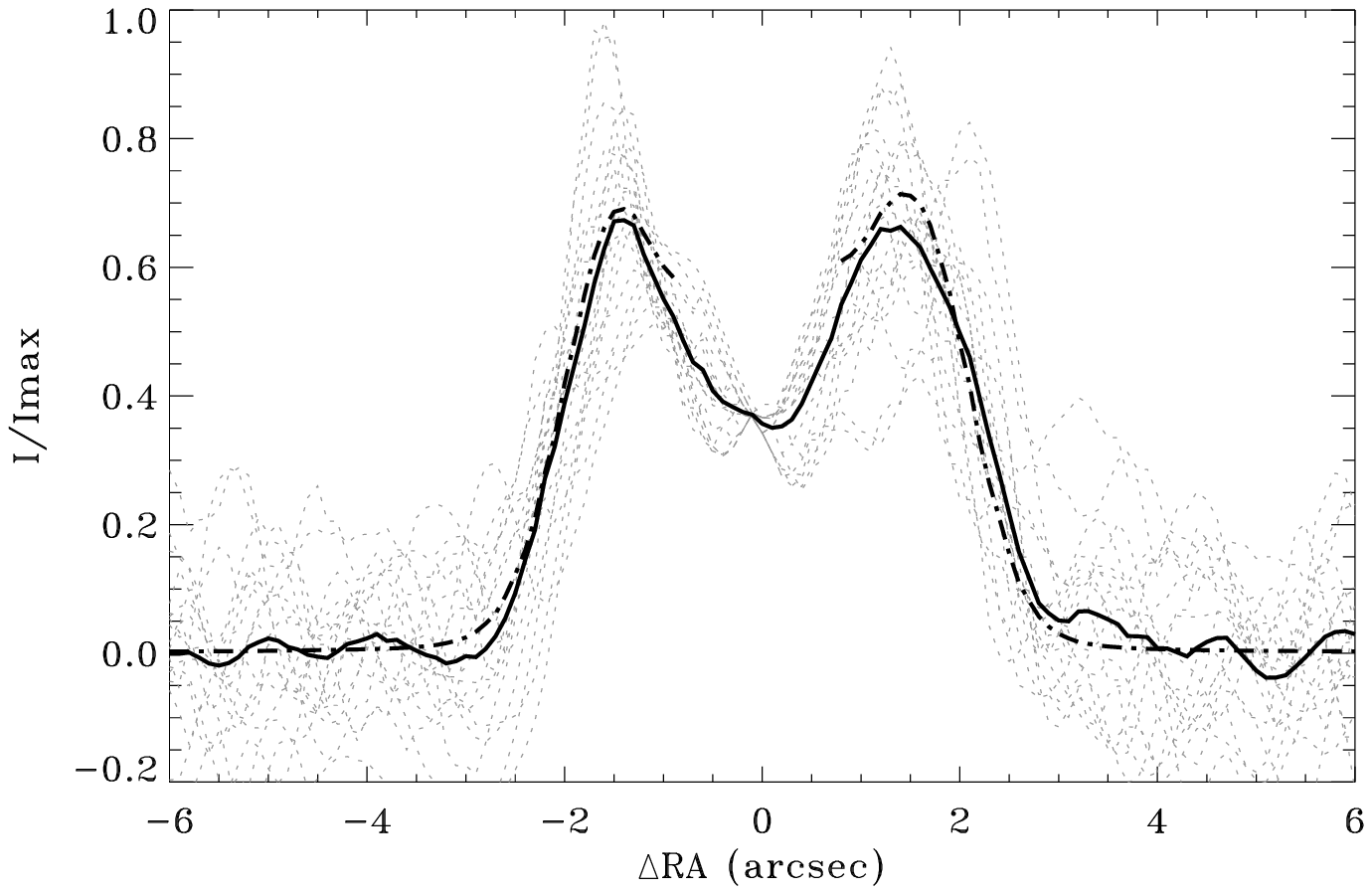}\end{center}
    \caption{Upper-left panel: map of S61 at \hbox{17\,GHz} obtained
      with the ATCA. The ellipse in the lower-left corner visualises
      the synthesised beam. Upper-right panel: archival
      $\rm H \rm \alpha$ {\it HST\/} image. Lower panel: surface
      brightness profiles. Grey dotted lines: profiles extracted
      from 18 cuts across the radio nebula, passing through the
      centre, in steps of $10^{\circ}$. Black line: mean of the grey
      dotted lines. Black dash-dotted line: mean surface brightness of the
      $\rm H \rm \alpha$ image, derived from 18 cuts as explained
      before.}
    \label{fig:radiomapslmc1}
  \end{minipage}
\end{figure}

\begin{table}
  \begin{center}
    \caption{Properties of the ATCA and ALMA maps.}
    \begin{tabular}{lcccccc}
      \hline
      Array&$\nu$ & HPBW  & LAS & PA & Peak & RMS\\
      \small
      &$(\rm GHz)$ & $('')$ & $('')$ & $(\rm deg)$ & \multicolumn{2}{c}{$(\rm mJy\, beam^{-1})$}\\
      \hline
      ATCA&17  & 0.836$\times$0.686 & 6.5 & -10.6           & 0.142 & 0.016\\
      ATCA&23  & 0.628$\times$0.514 & 4.1 & -10.8           & 0.121 & 0.032\\
      ALMA&343 &  1.23$\times$0.95  & 6.7 & \phantom{-}78.6 & 0.290 & 0.072\\
      \hline
    \end{tabular}
    \label{tab:ATCA2maps}
    \\
  \end{center}
\end{table}
We performed ATCA observations of S61 (together with two other Magellanic LBVs) between January 20 and 23, 2012. We used the array in the most extended configuration (6~km) and the Compact Array Broadband Backend (CABB) ``15~mm'' receiver in continuum mode. We split the 
receiver bandwidth in two 2-GHz sub-bands, one centred at
\hbox{17\,GHz} and the other at \hbox{23\,GHz}. This set-up was chosen in order 
to achieve enough spatial resolution to isolate possible contribution
from the central source and also to obtain some spectral information. 
We observed the phase calibrator ICRF~J052930.0$-$724528 for 1 minute,
alternating with 7- or 10-min scans on target, depending on the weather. For the bandpass correction we performed observations on the calibrator QSO~J1924$-$2914 each day as well as observations of the flux calibrator ICRF~J193925.0$-$634245. We also 
performed reference pointing frequently (about every 1--2 h) to
assure the pointing accuracy of the source observations. The total integration time obtained on each source was 8 hours.

We performed the data reduction and imaging using the MIRIAD package
\citep{1995SaultMiriad}. We split the datasets in two parts (one per
central frequency) and reduced them separately. For the data editing,
flagging and calibration, we followed the standard calibration recipe
for the millimetric band. We applied the opacity correction and
flagged bad data, before calculating corrections for gains. We used
observations of QSO J1924$-$2914, ICRF J193925.0$-$634245 and ICRF
J052930.0$-$724528 for determining the bandpass, flux density and
complex gain solutions, respectively. Once corrected, the visibilities
were inverted by Fourier transform. We chose the natural weighting
scheme of the visibilities, for best sensitivity. Deconvolution of the
dirty images was performed using the Clark algorithm \citep{1980Clark}
and the selection of the clean components was done interactively. We
then restored the clean components with the synthesised
beam. Table~\ref{tab:ATCA2maps} contains information about the
synthesised beam (Half Power Beam Width, HPBW) and position angle
(PA), largest angular scale (LAS), peak flux densities and rms-noise
of the resulting images. At \hbox{17\,GHz} we detect above $3\sigma$
the nebular emission in its whole extension. At \hbox{23\,GHz} the map
is noisy because of the system response to bad weather at higher
frequencies. For this reason we do not show the \hbox{23\,GHz}
data. The radio map at \hbox{17\,GHz} is illustrated in the upper-left
panel of Fig.~\ref{fig:radiomapslmc1}. 

We also include in our analysis the 5.5 and 9-GHz data from the ATCA observations performed in 2011 by means of the
CABB ``4cm-Band'' (4-10.8 GHz) receiver. These data
were presented in Paper I.

\subsection{The ALMA observation and data reduction}
\begin{figure}
  \centering
  \begin{minipage}{1\linewidth}
    \includegraphics[width=\textwidth]{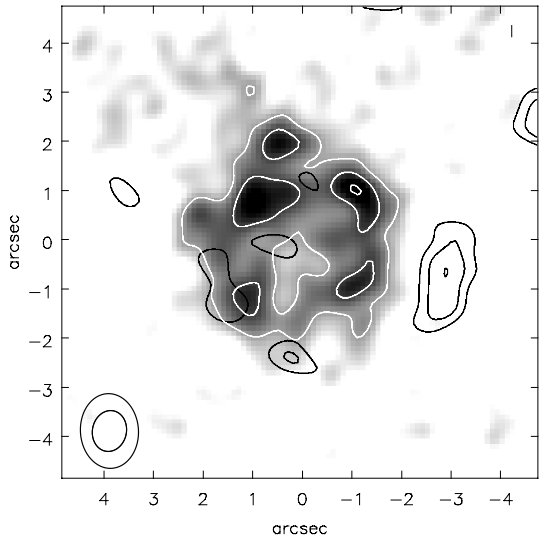}
    \caption{Raster ATCA \hbox{17\,GHz} map with $3, 7\sigma$ contours
      overlaid (white). Black contours: from 2 to 4$\sigma$ levels in
      the ALMA map at \hbox{343\,GHz}. The big and small ellipses in
      the left bottom corner are respectively the ALMA and ATCA
      synthesised beams.}
    \label{fig:alma}
  \end{minipage}
\end{figure}

S61 was observed as part of an ALMA Cycle-2 project studying three
Magellanic LBVs (2013.1.00450.S, PI Agliozzo). A single execution of 80 minutes total duration, including the three targets, was performed on 2014-12-26 with 40 12m antennas, with projected baselines from 10 to 245 m, and integration time per target of 16 minutes. A standard Band 7 continuum spectral setup was used, giving four 2-GHz width spectral windows of 128 channels of XX and YY polarisation correlations centred at approximately 336.5 (LSB), 338.5 (LSB), 348.5 (USB) and 350.5 (USB) GHz. Online, antenna focus was calibrated during an immediately preceding execution, and antenna pointing was calibrated on each calibrator source during the execution (all using Band 7). Scans at the science target tuning on bright quasar calibrators \bpcal and \fluxcal (PKS J0519-4546; an ALMA secondary flux calibrator `grid' source) were used for interferometric bandpass and absolute flux scale calibration. Astronomical calibration of complex gain variation was made using scans on quasar calibrator \qso, interleaved with scans on the science targets approximately every six minutes. Of the 40 antennas in the array, 36 were fully used in the final reduction, with two more partially used due to issues in a subset of basebands and polarisations. Data were calibrated and imaged with the Common Astronomy Software Applications (CASA) package \citep{CASA}.

Atmospheric conditions were marginal for the combination of frequency
and necessarily high airmass (transit elevation $45^\circ$ for
S61). Extra non-standard calibration steps were required to minimise
image degradation due to phase smearing, to provide correct flux
calibration, and to maximise sensitivity by allowing inclusion of
shadowed antennas. As S61 was not detected, we defer discussion of
these techniques to an article on the other sources in the sample
(Agliozzo et al. 2016, submitted).

We derived the intensity image from naturally weighted visibilities to
maximise sensitivity and image quality (minimise the impact of phase
errors on the longer baselines). We imaged all spectral windows
together ($343.5\,{\rm GHz}$ average; approximately $7.5\,{\rm GHz}$
usable bandwidth), yielding RMS noise of
$72\,\mu{\rm Jy}\,{\rm beam^{-1}}$ in the image. This is compared to
the proposed sensitivity of $40\,\mu{\rm Jy}\,{\rm beam^{-1}}$, which
could not be achieved as no further executions were possible during
the appropriate array configuration in Cycles 2 and 3. With this
sensitivity, we did not detect the nebula. In Fig.~\ref{fig:alma} we
show the 2, 3, and 4$\sigma$ contours in the ALMA map (in black) on
top of the ATCA \hbox{17\,GHz} image. These contours do not have
enough statistical significance. However, the elongated object West of
the radio nebula has a peak at $4\sigma$, but it is difficult to
associate it with S61. Details of the ALMA map are listed in
Table~\ref{tab:ATCA2maps}. Deeper observations with ALMA may detect
the nebular dust, and would certainly improve the constraints on the
dust mass. This would be a good candidate for the potential ``high
sensitivity array'' mode, combining all operational array elements
($12$ and $7\,{\rm m}$ antennas, typically at least 50 in total) in a
single array with the 64-input Baseline correlator, when in the more
compact 12m array configurations (this may be offered from Cycle 6 in
2018).

\subsection{VLT/VISIR observations}

\begin{table}
  \centering
  \caption{VISIR observational summary.}
  \label{tab:obsVISIR}
  \begin{tabular}{cccccc}
    \hline
    Date & Filter & Airmass & DIMM Seeing   & PWV\\
        &  &         & (arcsec) & (mm)\\
    \hline
   2015-09-03  &PAH2$\_$2 & 1.576 & 1.35 & 3.2\\   
  2015-09-04   &Q1        & 1.590 & 1.38 & 1.8\\
    \hline
  \end{tabular}
\end{table}

We proposed service-mode observations in the narrow bandwidth filters PAH2$\_$2 and Q1, centred
respectively at 11.88 and 17.65 $\rm \mu$m. The observations were
carried out between September 3rd and 4th, 2015. The observing mode was set for regular imaging, with
pixel scale of 0.045 arcsec. The OBs were executed in conditions slightly worse
(10\%) than specified in the scheduling constraints. Table
\ref{tab:obsVISIR} contains a summary of the VISIR observations.

We have reduced the raw data by running the recipe \texttt{$visir\_image\_combine.xml$} of the VISIR pipeline kit (version 4.0.7) in the environment Esoreflex 2.8. We have compared the calibrator (HD026967 and HD012524) flux densities and sensitivities with the ones provided by the observatory from the same nights, and have found consistent results. 
Due to the non-detection of the science target, the data reduction
pipeline has performed a straight combining of the images while
correcting for jitter information from the fits headers, rather than
stacking individual images with the shift-and-add strategy. In the last
step, the pipeline has converted the final (combined) images from ADU to $\rm Jy\,pixel^{-1}$ by
adopting the conversion factor derived from the calibrators. The
output of the pipeline is a single image of 851$\times$851 and
851$\times$508 pixels in the PAH2$\_$2 and Q1 filters, respectively. The two images are in
unit of $\rm Jy\,pixel^{-1}$. The rms-noise in the images is
0.08 $\rm mJy\,pixel^{-1}$ and 2.1 $\rm mJy\,pixel^{-1}$, in the
filters PAH2$\_$2 and Q1 respectively, which translate in noise of
$\sim 40$ and $\sim 1000\, \rm  mJy\,arcsec^{-2}$. We did not achieve
the expected sensitivity (as estimated with the Exposure Time
Calculator). This could be due to large scale emission. 
We estimate that to detect a point-like source with a signal-to-noise ratio
of 3, it should be at least 15 $\rm mJy$ at 11.88 $\rm \mu$m and 430
$\rm mJy$ at 17.65 $\rm \mu$m.

\subsection{Optical data}
The $\rm H \rm \alpha$ {\it Hubble Space Telescope\/} ({\it HST\/})
data \citep{Weis2003} where retrieved from the STScI data archive
(proposal ID: 6540), as already described in Paper I. They were
obtained with the Wide Field and Planetary Camera 2 (WFPC2) instrument
using the $\rm H \rm \alpha$-equivalent filter F656N and reduced by
the standard {\it HST\/} pipeline.  We combined the dataset (four
images with a 500~s exposure) following a standard procedure in IRAF
to remove cosmic-ray artefacts and to improve the signal-to-noise
ratio SNR.  We also recalibrated the {\it HST\/} image astrometrically
using the Naval Observatory Merged Astrometric Dataset (NOMAD)
catalogue \citep{2005Zac} for a corrected overlay with the radio
images.  Finally, we converted the {\it HST\/}/WFPC2 image from
$\rm counts pixel^{-1}\,\AA^{-1}$ units to $\rm erg\,cm^{-2}\,s^{-1}$
units, by multiplying with $2.9\times10^{-16}\times21.5$ (where 21.5
is the F656N filter bandwidth in $\rm \AA$).

\section{The Radio Emission}
\label{sec:analysis}

\begin{figure}
  \centering
  \begin{minipage}{1\linewidth}
    \includegraphics[width=1\textwidth]{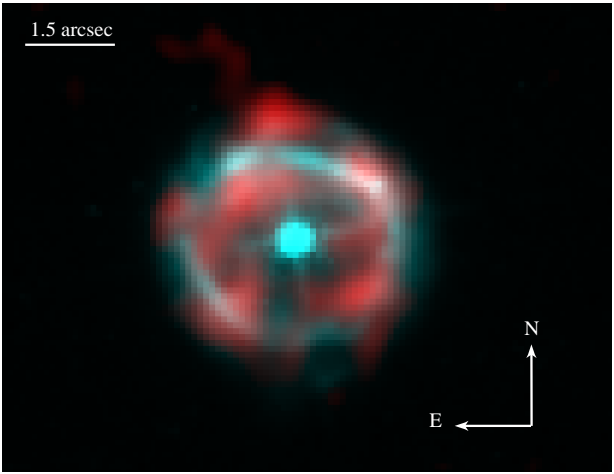}
    \caption{Two-colour image of S61. The red colour is the radio \hbox{17\,GHz}, the cyan one is the $\rm H \rm \alpha$.}
    \label{fig:rgb}
  \end{minipage}
\end{figure}

\begin{table}
  \begin{center}
    \caption{Observed flux densities, angular sizes and spectral index}
    \label{tab:fluxobs2}
    \begin{tabular}{ccccc}
      \hline
      S(\hbox{5.5\,GHz})&S(\hbox{9\,GHz})&S(\hbox{17\,GHz})& Size& $\langle\alpha\rangle$\\
      (mJy)  &(mJy) &  (mJy)     & (arcsec)&\\
      \hline
      2.1$\pm$0.1 & 2.2$\pm$0.3 & 1.97$\pm$0.10 & 4.5$\times$4.9 & -0.06$\pm$0.06\\ 
      \hline
    \end{tabular}
    \footnotesize
  \end{center}
\end{table}
\begin{figure*}

 \begin{minipage}{1\textwidth}
\centering

\includegraphics[scale=.3]{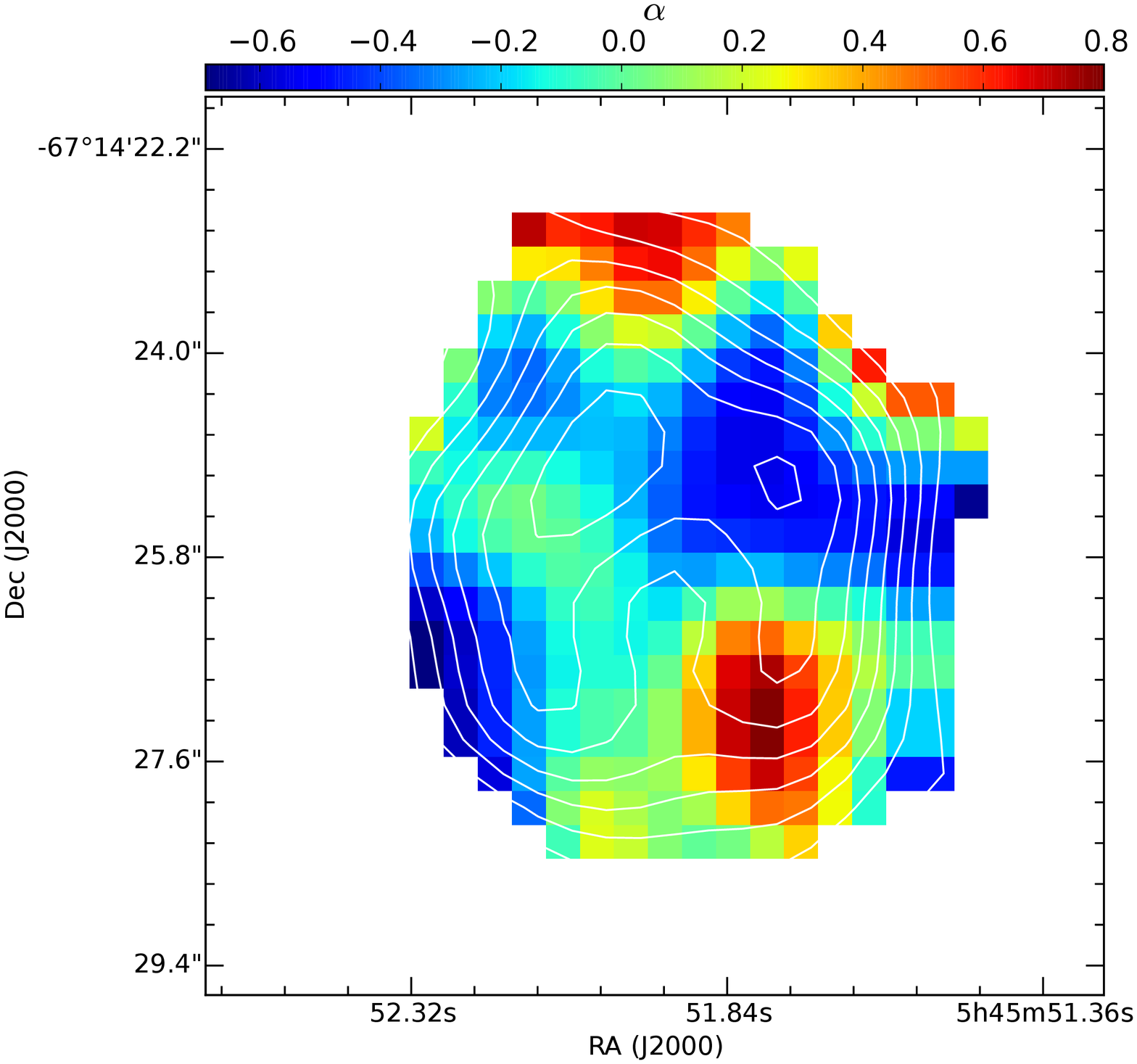}
 \includegraphics[scale=.3]{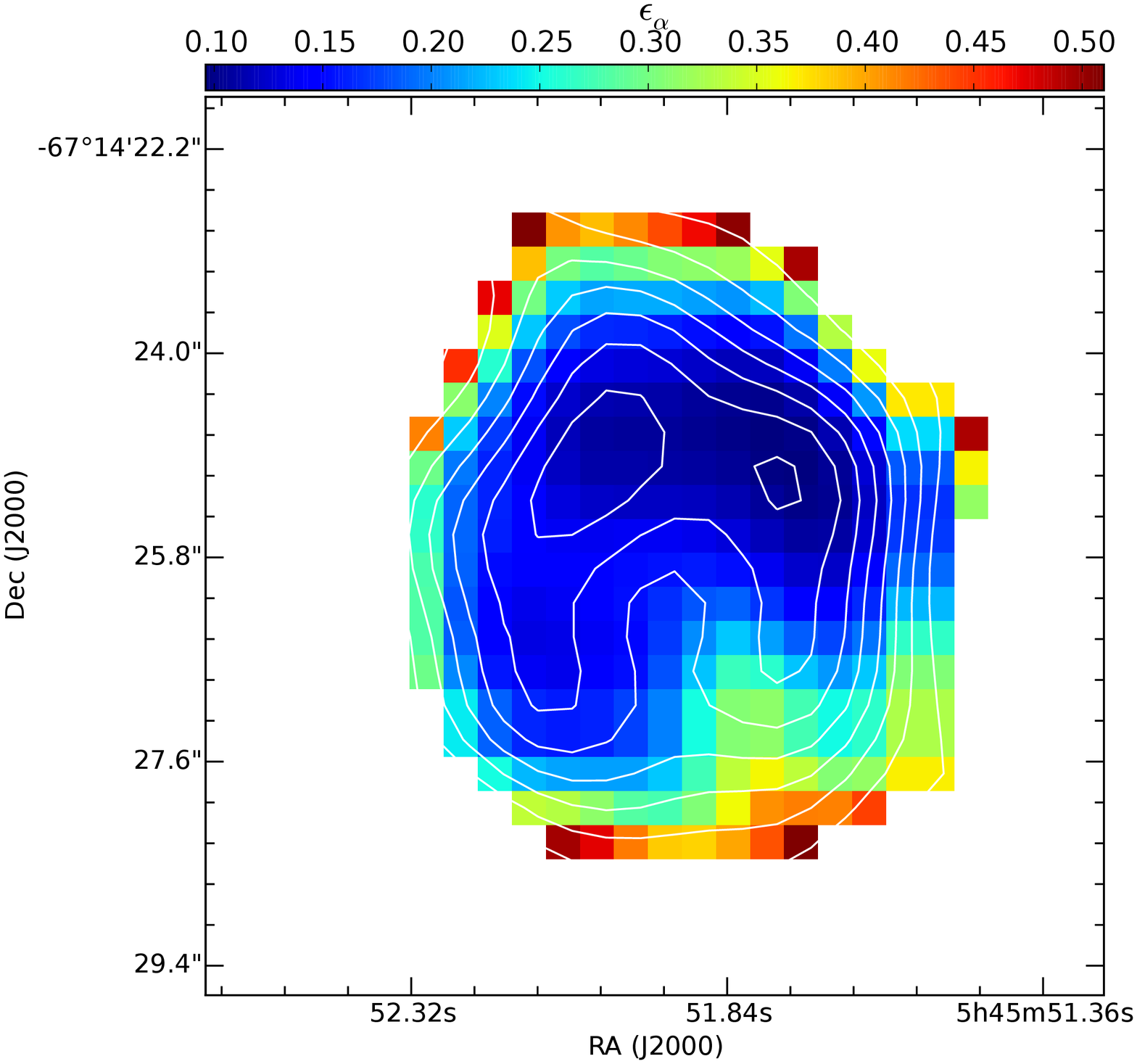}
 \caption{Left: spectral index map between 9 and \hbox{17\,GHz}.
   Right: spectral index error map. Before the computation, the map at
   \hbox{17\,GHz} was reconvolved to match the beam at \hbox{9\,GHz}. The white
   contours indicate flux densities above 3 $\sigma$ (at integer
   steps) of the \hbox{17\,GHz} emission at the \hbox{9\,GHz} resolution, and the
   cross represents the position of the star.}
\label{fig:spix_lbvs}
 \end{minipage}
\end{figure*}
In the upper panel of Fig.~\ref{fig:radiomapslmc1}, the radio map
(left) is compared with the $\rm H \rm \alpha$ {\it HST\/} image
(right). For a better visualisation of the nebular morphology, we also
show in the lower panel the radial surface brightness profiles,
extracted from 18 cuts (grey dotted lines) across the radio
nebula, passing through the centre and successively rotated by
$10^{\circ}$.  The black line is the arithmetic mean of the grey
dotted lines. In a similar way we derived the mean surface brightness
of the $\rm H \rm \alpha$ {\it HST\/} image (black dash-dotted line), after
convolving it with the radio beam. To block the emission of the
central object in the $\rm H \rm \alpha$ image, we applied a mask at
the position of the star. The image shows that there is more
substructure in the radio than in the optical, as is clearly evident
by comparing the surface profiles.  At \hbox{17\,GHz} the nebula size
is similar to the one in the $\rm H \rm \alpha$ image. A two-colour
image of the \hbox{17\,GHz} and $\rm H \rm \alpha$ data is shown in
Fig.~\ref{fig:rgb}. In the Northern part, apparently attached to the
shell, there is a spur-emission, similar to G79.29+0.46
\citep[e.g.][]{1994Higgs}. This compact object does not have a
counterpart either at lower frequencies or in the optical. It might
indicate an optically thick medium at the radio wavelengths.

\citet{Weis2003} reported that the optical nebula of S61 is expanding spherically and with a velocity of $\sim\!27\ \rm km\,s^{-1}$,
although slightly red-shifted to the West and blue-shifted to the
East, which they ascribed to a geometric distortion along the line of
sight. The radio nebula at \hbox{17\,GHz} is consistent with the shell geometry and therefore we will take it into account to model the radio emission (Section \ref{sec:model}). 

Table~\ref{tab:fluxobs2} lists the spatially integrated flux density and its associated
error at \hbox{17\,GHz}, together with the estimated nebula angular sizes (not
deconvolved by the synthesised beam).  The integrated flux density was determined by using the CASA
\texttt{viewer}. In particular, we selected with the polygonal tool
the area above $3\sigma$ level and integrated the emission over the
nebula. The rms-noise in the map was evaluated in regions free of
emission and hence flux density errors were estimated as
$\epsilon=\sigma \sqrt{N}$, where $N$ is the number of independent
beams in the selected region. Calibration flux density errors are usually negligible at these frequencies. The flux densities at 5.5 and \hbox{9\,GHz} derived in Paper I are also shown in the table.

From analysis of the spectral index, we can obtain information about the nature of the radio emission. We have computed the mean spectral index $\langle\alpha\rangle$ through a weighted fit of the power-law $S_{\nu}\propto\nu^{\alpha}$ between the flux densities at 5.5, 9 and \hbox{17\,GHz} in Table~\ref{tab:fluxobs2}. The ``global'' spectral index $\langle\alpha\rangle=-0.06\pm0.06$ is consistent with optically  thin free-free emission. 
 We have also obtained a spectral index map (per-pixel). To this end,
 the highest-resolution map (\hbox{17\,GHz}) was re-gridded and
 convolved with the beam at lower frequency. We show the spectral index
 map between \hbox{17\,GHz} and \hbox{9\,GHz} (with the beam about $1.5\times1.2$
 arcsec$^{2}$, as in Paper I) and its associated error map in
 Fig.~\ref{fig:spix_lbvs}. Since calibration errors are negligible, in
 the maps the error in each pixel is mostly given by the sum in
 quadrature of the rms-noise in both the maps, in Jy pixel$^{-1}$ units. 
In the inner part of the nebula the mean spectral index is \hbox{$\langle\alpha\rangle=-0.3\pm0.2$} (where the error is the mean value in the error map), consistent with optically thin free-free emission. In the Southern (bottom) part we observe a higher spectral index ($\alpha_{max}=0.8\pm0.3$) suggesting some mechanism of self-absorption of the free-free emission, due to, for instance, density clumps. The spectral index analysis may be biased by the fact that the interferometer is sensitive to different large and intermediate angular scales at different frequencies. However, the largest angular scales covered in the two datasets (12.9 and 6.5 arcsec at 9 and \hbox{17\,GHz}, respectively) are larger than the size of the nebula (Table~\ref{tab:fluxobs2}). We also rely on a good \emph{uv} coverage at the intermediate angular scales acquired during the observations. 

\section{Modelling the nebula}
\label{sec:model}

In Paper I we derived an estimate of the ionised mass in the S61 nebula from the
\hbox{9\,GHz} ATCA map.  Simply, the total ionised mass can be estimated if the density of particles and the volume of the nebula are known. For
non-self-absorbed optically thin free-free emission, the electron density $n_e$ 
can be determined through the relation between the emission measure,
\begin{equation}
  EM = \int_{0}^{s}{n_e^2\ \dif s\ [\rm pc\,cm^{-6}]},
  \label{equ:emmeasure}
\end{equation}
and the optical depth $\tau_{ff}(\nu)$ at frequency $\nu$
\begin{equation}
  \tau_{ff}(\nu) = 8.24\times10^{-2}\left(\frac{T_{e}}{\rm K}\right)^{-1.35}\left(\frac{\nu}{\rm 5\,GHz}\right)^{-2.1}\frac{EM}{\rm pc\,cm^{-6}}.
\end{equation}
$\tau_{ff}(\nu)$ can be determined from the
solution of the radiative transfer equation ($B_{\nu}=B_{bb}(T) \tau_{ff}(\nu)$) by setting as $B_{\nu}$ the radio brightness and by assuming a blackbody
with temperature $T$ equal to the electron temperature $T_e$.  Therefore, in Paper I we derived an average $n_e$ from the mean $EM$ (integrated over the nebula) and assumed as $s$ the transversal size of the nebula (measured on the radio map). With these values, we estimated for S61's nebula an ionised mass of $\sim 0.8\,\rm \Msun$. In reality, $n_e$ may vary inside the nebula. Furthermore, the geometrical depth $s$ may vary for different line of sights and then requires a proper geometrical model. 
Therefore, we propose a new approach
to fit all the pixels of the radio maps with a global
geometrical 3D density model of the nebula. Obviously, the nebula has to be spatially resolved. Instrumental effects on the nebula size due to bad resolution have to be negligible. If not, the estimated mass may be inaccurate.  

\subsection{3D density model \rhocube}
\label{sec:rhocube}

We have written and make publicly
available\footnote{\url{https://github.com/rnikutta/rhocube}}
\rhocube\ \citep{rhocube}, a \textsc{Python} code to model 3D density
distributions \rhoxyz\ on a discrete Cartesian grid, and their
integrated 2D maps \hbox{$\int\dif z \rhoxyz$}.
It can be used for a range of applications; here we model with it the
electron number density $n_e(x,y,z)$ in LBV shells, and from it
compute the emission measure $EM$ given in Eq.~\eqref{equ:emmeasure}.

The code repository includes several useful 3D density distributions,
implemented as simple \textsc{Python} classes, e.g. a power-law shell,
a truncated Gaussian shell, a constant-density torus, dual cones, and
also classes for spiralling helical tubes.
Other distributions can be easily added by the user.
Convenient methods for shifts and rotations in 3D are also provided.
If necessary, an arbitrary number of density distributions can be
combined into the same model cube, and the integration $\int\dif z$
will be correctly performed through the joint density field.
Please see Appendix \ref{sec:own-density} and the code repository for
usage examples of \rhocube, and for details of the
implementation.

\subsection{Bayesian parameter inference}
\label{sec:bayesian_inference}

We will apply \rhocube\ to our problem of estimating the physical
parameters of the observed emission maps by modelling the underlying 3D
electron density distribution $n_e(x,y,z)$.
We employ a Bayesian approach and compute marginalized posterior
density distributions of model parameters from the converged chains of
Markov-Chain Monte Carlo (MCMC) runs.
\emph{Bayes' Theorem} and the details of MCMC sampling are described
in Appendix \ref{sec:bayparinf}.
Our routines to fit \rhocube\ models to data are available as
supplemental materials, in Section \ref{sec:supplements}.

We begin with any 3D model for geometry of the nebula, e.g. a
truncated Gaussian shell. At every sampling step of the MCMC procedure
we draw a random vector $\theta$ of values for the free model
parameters (e.g. the shell radius, width, lower and upper truncation
radii, and $x$ and $y$ offsets). We then compute the 3D electron
density distribution $n_e(x,y,z|\theta)$, and from it the squared
integrated 2D map $EM_{\rm mod} = \int\! n_e^2(x,y)\, \dif z$.
The normalisation of $n_e$ is at first arbitrary, but by comparison
with the observed map $EM_{\rm obs}$ we can find a global scale $S$
such that the likelihood is maximised, or equivalently, the
chi-squared statistic
\hbox{$\chi^2 = \sum_i (d_i - S \cdot m_i)^2 / \sigma_i^2$} is
minimised.
The $d_i$ are the measured pixel values $EM_{\rm obs}^{\,i}$ (for unmasked
pixels only), and $m_i$ their modelled counterparts.
$S$ can be computed analytically \citep[e.g.][]{Nikutta2012phd}. Note
that the pixels $i$ are \emph{independent}.

\subsubsection{Application to the S61 nebula}
\label{sec:rhocube_application_S61}
\begin{table}
  \center
  \begin{tabular}{lcrr}
    \hline
    Parameter            & Units         &     \hbox{9\,GHz}  & \hbox{17\,GHz}          \\
    \hline
    Truncated Normal Shell & & & \\[3pt]
    $r$                  &pc        & $[0, 0.3]$       &  $[0, 0.3]$     	\\[3pt]
    $\sigma_r$      	 &pc	    & $[0, 0.7-r]$     &  $[0, 0.6-r]$   	\\[3pt]
    $r_{\rm lo}$     	 &pc	    & $[0, r]$         &  $[0, r]$       	\\[3pt]
    $r_{\rm up}$    	 &pc	    & $[0, r+\sigma_r]$&  $[0, r+\sigma_r]$	\\[3pt]
    Power-Law Shell & & & \\[3pt] 
    $r_{in}$     	 &pc	    & $[0, 0.3]$              &  $[0, 0.3]$           	\\[3pt]
    $r_{out}$ 	    	 &pc        & $[r_{in}, 0.7]$              &  $[r_{out}, 0.6]$\\          	
    \hline
  \end{tabular}

  \caption{Uniform priors adopted for the truncated normal shell and
    the power-law shell models. The priors were chosen from visual
    inspection of the 9 GHz and 17 GHz maps and were limited to
    meaningful ranges.}
  \label{tab:priors}
\end{table}

\begin{table}
  \center
  \begin{tabular}{lcrrrr}
    \hline
    Parameter                 & Units         & \multicolumn{4}{c}{Truncated Normal Shell}                      \\

                              &     	      & \multicolumn{2}{c}{MAP} & \multicolumn{2}{c}{Median}                         \\
                              &     	      &  \hbox{9\,GHz} & \hbox{17\,GHz} & \hbox{9\,GHz} & \hbox{17\,GHz}                  \\
                              &     	      &  ($\chi^2_{r} = 46$) & ($\chi^2_{r} = 33$)        & &               \\

    \hline
    $r$              		&pc	&  0.19  &   0.30      &  $ 0.19^{+0.07}_{-0.09}$ & $ 0.18^{+0.08}_{-0.09}$\\[3pt]
    $\sigma_r$     	  	&pc	&  0.50  &   0.30      &  $ 0.35^{+0.13}_{-0.15}$ & $ 0.29^{+0.12}_{-0.13}$\\[3pt]
    $r_{\rm lo}$  		&pc	&  0.05  &   0.28      &  $ 0.08^{+0.09}_{-0.06}$ & $ 0.07^{+0.09}_{-0.05}$\\[3pt]
    $r_{\rm up}$  		&pc	&  0.67  &   0.58      &  $ 0.35^{+0.15}_{-0.12}$ & $ 0.31^{+0.13}_{-0.12}$\\[3pt]
    x-offset        		&pc     & -0.07  &  -0.04      &  $ 0.00^{+0.09}_{-0.10}$ & $ 0.00^{+0.03}_{-0.03}$\\[3pt]
    y-offset          		&pc     & -0.08  &  -0.01      &  $ 0.00^{+0.09}_{-0.10}$ & $-0.00^{+0.03}_{-0.03}$\\[3pt]
    M$_{\rm ion}$ 		&\Msun\ &  0.29  &   0.13      &  $ 0.07^{+0.09}_{-0.04}$ & $ 0.03^{+0.04}_{-0.02}$\\
    \hline
  \end{tabular}
  \caption{Inference of model parameters, as derived from data at \hbox{9\,GHz} and \hbox{17\,GHz}. A truncated
    normal spherical shell was used as a model. MAP =
    maximum-a-posteriori values. Median = median of marginalized posterior distributions,
    with $1\sigma$ confidence intervals. Note that M$_{\rm ion}$ is a
    derived quantity, i.e. not a free (modelled) parameter.}
  \label{tab:S61statsTNS}
\end{table}

\begin{landscape}
\begin{table}
  \center
  \begin{tabular}{lcrrrrrrrrrrrrrrrr}
   \hline 
   Parameter              & Units & 		\multicolumn{4}{c}{PLS exp=0 (constant-density shell)}		    &	\multicolumn{4}{c}{PLS exp-1}                        &	\multicolumn{4}{c}{PLS exp=-2}\\
			  &       &\multicolumn{2}{c}{MAP} 	& \multicolumn{2}{c}{Median}        & \multicolumn{2}{c}{MAP}    & \multicolumn{2}{c}{Median}        & \multicolumn{2}{c}{MAP}    & \multicolumn{2}{c}{Median} \\

                          &       &\hbox{9\,GHz}&\hbox{17\,GHz} &\hbox{9\,GHz} &\hbox{17\,GHz}        &\hbox{9\,GHz} &\hbox{17\,GHz} &\hbox{9\,GHz}&\hbox{17\,GHz}       &\hbox{9\,GHz}&\hbox{17\,GHz} &\hbox{9\,GHz}&\hbox{17\,GHz}\\
                           &       &($\chi^2_{r} = 47$)&($\chi^2_{r} = 22$)&& &($\chi^2_{r} = 55$)&($\chi^2_{r} = 48$)&& &($\chi^2_{r} = 100$)&($\chi^2_{r} = 74$)&&\\
   \hline
   $r_{in}$           &pc   & 0.14  &  0.26  &$ 0.16^{+0.09}_{-0.08}$&$ 0.17^{+0.08}_{-0.10}$  & 0.26 &  0.28  &$ 0.16^{+0.09}_{-0.08}$& $ 0.17^{+0.08}_{-0.08}$ & 0.29 &  0.30  &$ 0.16^{+0.09}_{-0.08}$&$ 0.16^{+0.09}_{-0.07}$\\[3pt]
   $r_{out}$          &pc   & 0.69  &  0.59  &$ 0.45^{+0.17}_{-0.18}$&$ 0.39^{+0.16}_{-0.15}$  & 0.69 &  0.57  &$ 0.44^{+0.17}_{-0.18}$& $ 0.39^{+0.15}_{-0.12}$  & 0.69 &  0.59  &$ 0.44^{+0.17}_{-0.18}$&$ 0.37^{+0.16}_{-0.15}$\\[3pt]
   x-offset           &pc   &-0.02  & -0.02  &$-0.00^{+0.02}_{-0.02}$&$ 0.00^{+0.01}_{-0.02}$  &-0.03 & -0.02  &$ 0.00^{+0.02}_{-0.02}$& $-0.00^{+0.02}_{-0.02}$ & -0.03& -0.02  &$ -0.00^{+0.02}_{-0.02}$&$ 0.00^{+0.01}_{-0.02}$\\[3pt]
   y-offset           &pc   &-0.03  &  0.00  &$ 0.00^{+0.02}_{-0.02}$&$ 0.00^{+0.02}_{-0.02}$  &-0.03 & -0.02  &$ 0.00^{+0.02}_{-0.02}$& $-0.00^{+0.02}_{-0.02}$ & -0.03& -0.00   &$-0.02^{+0.02}_{-0.02}$&$ 0.00^{+0.02}_{-0.01}$\\[3pt]
   M$_{\rm ion}$      &\Msun&0.31&  0.14  &$ 0.12^{+0.12}_{-0.09}$&$ 0.05^{+0.07}_{-0.04}$  & 0.30 &  0.12  &$ 0.11^{+0.12}_{-0.08}$& $ 0.04^{+0.05}_{-0.03}$ & 0.27 &  0.11  &$ 0.07^{+0.08}_{-0.05}$&$ 0.03^{+0.04}_{-0.02}$\\
    \hline
  \end{tabular}
  \caption{Inference of model parameters, in the case of power-law shells, with exponents: $0$ (i.e. constant-density shell), $-1$ and $-2$. MAP =
    maximum-a-posteriori values. Median = median of marginalized posterior distributions,
    with $1\sigma$ confidence intervals. Note that M$_{\rm ion}$ is a
    derived quantity, i.e. not a free (modelled) parameter.}
  \label{tab:S61statsPLS}
\end{table}
\end{landscape}
\begin{figure*}
  \centering

    \includegraphics[width=\textwidth]{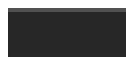}

  \includegraphics[width=\textwidth]{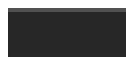}

  \caption{Results of fitting the \hbox{9\,GHz} (top two rows) and
    \hbox{17\,GHz} (bottom two rows) EM maps of S61 with truncated Gaussian
    shells. Panel (0) shows the data with linear sizes relative to the
    central pixel, with a $3\sigma$ mask applied. Panel (1) shows the
    model shell with parameters corresponding to the median values of
    all marginalized posterior distributions. Indicated as solid,
    dashed, and dotted circles are the median values of $r$,
    $r\pm\sigma_r$, and $r_{\rm up}$, respectively. Panels (2)-(6)
    show as red histogram the posteriors of four model parameters, and
    of the derived ionised mass M$_{\rm ion}$. All histograms are
    normalized to unit area. The smooth black curve is a Gaussian
    kernel density estimation. The vertical blue-dashed and
    green-dotted lines indicate the maximum-a-posteriori (MAP) and
    median value of the MCMC chain, respectively. The blue 
    dashed line in panel (6) locates the ionised mass value of
    the model shown in panel (1). Panel (7) illustrates azimuthal mean
    profiles derived from 18 extracted cuts along the EM (data: blue
    line; model: red line). }
  \label{fig:s61_posteriors}
\end{figure*}

In the following we apply \rhocube\ together with MCMC and Bayesian
inference to fit the EM maps obtained for S61 from the data at \hbox{9\,GHz}
and \hbox{17\,GHz}. Note that the angular resolution in the map at \hbox{9\,GHz} is poor and affects the nebula size (Paper I). We only use this data to test our procedure and to compare the mean $EM_{\rm mod}$ and the derived ionised mass with those estimated in Paper I. 

The radio maps of S61, while irregular, are indicative of a
spherical matter distribution (e.g. Fig.~\ref{fig:radiomapslmc1},
top-left).
This is even more apparent in the \Ha\ map
\citep[Fig.~\ref{fig:radiomapslmc1}, top-right; see also][]{1999Pasquali, Weis2003}.
We therefore modelled with \rhocube\ the electron density distribution
by using the following geometries: a \emph{truncated Gaussian (normal)
  shell} and a \emph{power-law shell}. For the \emph{truncated normal
  shell} (hereafter TNS) the free parameters are six: the shell radius
$r$, the width $\sigma_r$ of the Gaussian around $r$, the lower and
upper clip radii $r_{\rm lo}$ and $r_{\rm up}$, and finally we allow
for minute offsets in the plane of the sky, \xoff\ and \yoff, to
account for possible de-centring of the observed shell. For the
\emph{power-law shell} (PLS) the free parameters are four: the inner
and outer radii $r_{in}$ and $r_{out}$, \xoff\ and \yoff. For the PLS
geometry we explored the cases with exponents: 0 (i.e. constant
density shell), -1 and -2. We noticed that exponents smaller than -2
were producing lower-quality results and therefore we will not comment
on them.
We used uniform prior probability distributions (i.e. before
introducing the model to the observed data) on the shell radius,
width, and both clip radii for the TNS geometry, and for the inner and
outer radii for the PLS geometry.

The ranges of these parameters must of course be limited to meaningful
values. For convenience we converted the $x$ and $y$ pixel units in
the maps from arcsec to parsec, assuming a distance of 48.5
kpc. Hence, from visual inspection of the map we adopted the values
listed in Table~\ref{tab:priors}.  As priors of both offsets \xoff\
and \yoff\ we adopted very narrow Gaussians, truncated at
$\pm 2$~pixels from the central pixel. The maps supplied to the code
are $41\times41$ and $101\times101$ pixels (for the \hbox{9\,GHz} and
\hbox{17\,GHz} data, respectively). In Tables \ref{tab:S61statsTNS}
and \ref{tab:S61statsPLS} we show the resulting marginalized
posteriors from the geometries mentioned before.

Figure \ref{fig:s61_posteriors} illustrates only the posteriors from the fit of the \hbox{9\,GHz} and \hbox{17\,GHz} data with a truncated Gaussian shell. These posteriors were obtained after drawing $2\times10^4$~MCMC samples.  
Many fewer samples are necessary for convergence ($\sim$1000 may be
sufficient), but more samples produce smoother histograms.
In the figure we do not show the posteriors for the offsets \xoff\ and
\yoff\ which are very narrow and centred, i.e. the shell is not
significantly shifted from the central pixel. 
The MCMC chain histograms are shown in red, a Gaussian kernel density
estimation is overplotted in black.\footnote{Computed with
  \textsc{Seaborn}, available
  from\\\url{http://stanford.edu/~mwaskom/software/seaborn}}
Blue-dashed vertical lines indicate the single best-fit values
(maximum-a-posteriori, MAP) of the MCMC chains, i.e. the combination
of parameter values which simultaneously maximise the likelihood.
Note that this need not be the ``most typical'' solution.
Green-dotted lines mark the median values of the marginalized
posterior PDFs.
These statistics, and the $1\sigma$ confidence intervals around the
median, are summarised in Table~\ref{tab:S61statsTNS}.
Note that for the \hbox{9\,GHz} data PLS (with exp=0) equally produces a good-quality result.
For the \hbox{17\,GHz} data the PLS (with exp=0) produces the best-fit.  
The MAP models (for
unmasked pixels) have a formal $\chi^2_{r}$ as shown in the tables. 
While large values, considering the simple model and the clearly not
entirely spherical/symmetric EM map, they are acceptable.
In the figure we only show our favourite models, chosen because they produce more agreeable distributions and radial profiles. For this reason we now describe only them. 

The distribution of shell radii (panels 2 in
Fig.~\ref{fig:s61_posteriors}) is quite broad, but clearly peaks within the shell. 
The radial thickness of the Gaussian shell is symmetric around the peak in both panels 3. A similar comment can be given for the upper clip radius  $r_{\rm up}$ (panels 5).
The lower clip radius $r_{\rm lo}$ (panels 4) is
left-bounded at 0, and the decline at the large-values tail is driven by
our prior requirement.
The observed \hbox{9\,GHz} and \hbox{17\,GHz} $EM_{obs}$ maps are shown in panels (0), and the model shell corresponding to
the MAP model is shown in (1).
This panel also shows with solid, dashed and dotted circles the
median-model values of $r$, $r\pm\sigma_r$, and $r_{\rm up}$,
respectively. 
Panel (7) illustrates azimuthal mean profiles derived from 18 extracted cuts along $EM_{obs}$ (blue line) and $EM_{\rm MAP}$ (red line). While the MAP model for the \hbox{9\,GHz} seems underestimated, the \hbox{17\,GHz} model profile is quite satisfactory.

As evident in the tables, different models (TNS, PLS exp=0, PLS
exp=-1) produce similar-quality results, meaning that with the current
data we cannot constrain the electron density distribution in the S61
nebula.

\subsection{Ionised mass}

Knowing the 3D distribution of the electron number density
$n_e(x,y,z)$, we can now compute the total ionised mass contained in
the shell via
\begin{equation}
  {\rm M_{ion}} = \frac{m_p}{\Msun} \int\! \dif V\ n_e(x,y,z),
  \label{equ:mass}
\end{equation}
with $m_p$ and \Msun\ the proton and solar masses, under the
assumption that the gas comprises only ionised hydrogen.
For simple symmetric geometries and density distributions this
integral can be evaluated analytically (e.g. a constant-density
shell), but it might be significantly more challenging for more
complicated geometries and more complex density fields.
In our discretised 3D Cartesian grid, realizing that the volume of a
3D-voxel is $(\Delta x)^3$, because $\Delta x = \Delta y = \Delta z$,
Eq. \eqref{equ:mass} simplifies to
\begin{equation}
  \label{equ:mass_grid}
  {\rm M_{ion}} = \frac{m_p}{\Msun}\, (\Delta x)^3 \sum_i n_e(x_i,y_i,z_i),
\end{equation}
where the index $i$ runs over all voxels (recall they are
independent).

Thanks to the MCMC approach we can use the entire converged chains of
model parameter values to compute posterior distributions of
\emph{derived} quantities (i.e. not \emph{modelled} quantities), such
as the ionised mass here.
The resulting marginalized posterior distribution for the TNS geometry is shown in
panels (6) of Fig.~\ref{fig:s61_posteriors}, with the purpose to provide an example. In fact, as mentioned before, we do not have a statistically strong model to discern among the possible density distributions.  However, it is comforting to see that all the models  produce similar masses (see Tables \ref{tab:S61statsTNS} and \ref{tab:S61statsPLS}). 

The TNS model of the 9-GHz (17-GHz) data generates a peaked and skewed
distribution of $M_{ion}$, with median $0.07^{+0.09}_{-0.04}~\Msun$
($0.03^{+0.04}_{-0.02}~\Msun$).
The MAP-model values are $0.29~\Msun$ and $0.13~\Msun$. These are located always in the right side of the distribution, probably due to our prior requirements to fit the shell within the edge of the nebula, rather than the edge of the image frame. 
We remind the reader that the modelling of the 9 GHz data was proposed in order to test the code and to compare the results with our previous estimation \citep[$\sim$0.8~\Msun,][]{2012Agliozzo}. 
The value derived with this new approach is about 2.7 times smaller
than the previous estimate. However, because of the asymmetry of the
nebula at 9 GHz, the model seems to underestimate, on average,
$EM_{obs}$ (see Panel 7 in Fig. \ref{fig:s61_posteriors}). According
to this, the two methods may not disagree each other.The advantage of the proposed new method
is that it requires no assumptions about
the nebula depth $s$.

The derived mass from the fits of the 17 GHz data
($0.11-0.14\,\rm M_{\odot}$) are more representative, because of the
smaller $\chi^{2}_{r}$ than the 9 GHz data. The mean profile of
$EM_{MAP}$ reproduces satisfactorily $EM_{obs}$ (see bottom Panel 7 in
Fig. \ref{fig:s61_posteriors}). More importantly, the angular
resolution achieved at 9 GHz affects the nebular size, resulting in a
larger volume to model. For further analysis we will then adopt the
MAP model from the fit of the 17 GHz data.
  
The mass estimated here are at least an order of magnitude
smaller than the one derived by \citet{1999Pasquali} from the
$ \rm H \rm \alpha$ luminosity and from optical emission lines. The
discrepancy may be due to a combination of different assumptions and
methods. For instance, \citet{1999Pasquali} assumed for the LMC a
distance of 51.2 kpc and measured a size for the nebula of 7.3 arcsec
(i.e., 1.8 pc) from an image with a poorer resolution. Their
estimation of the average electron density was also uncertain due to
uncertainties of the [SII] 6717/6731 ratio. It would be interesting to
compare our results with integral field unit observations of the
nebula around S61.

As shown in Section \ref{sec:analysis}, there are hints of inhomogeneities in the nebula. Following \citet{1981Abbott}, who described the radio spectrum of a clumped stellar wind, we can assume discrete gas clumps of relatively higher
density ($n_{H}$), embedded in a lower-density medium ($n_{L}$). Both the
clumps and the inter-clump medium are assumed optically thin. The clumps
are distributed randomly throughout the volume of the nebula. If we define $f$ as the fractional volume which contains material at density $n_{H}$, then Eq. (\ref{equ:emmeasure}) becomes

\begin{equation}
EM = \int_{0}^{s}{(f n_H^2 + (1-f) n_L^2)\ \dif s\ }.
\end{equation}

The ionised mass in the nebula can be underestimated if not corrected for clumpiness. In fact, in the simple case of an empty inter-clump medium ($n_L=0$), it is possible to demonstrate that $M\propto S_{\nu}^{\frac{1}{2}} f^{-\frac{1}{2}}$. For a filling factor $f=0.5$ the ionised mass would be about 41$\%$ larger than in the case of a homogeneous nebula. For a more generic case, the factor to correct the estimated ionised mass is 
 $(f n_H^2 + (1-f) n_L^2)^{\frac{1}{2}}$. The ionised mass could also be underestimated in the emitting regions with a positive spectral index. In the specific case of S61, where $\alpha\sim0.8$, the optical depth is $\lesssim1$. Moreover, the region with positive $\alpha$ is small. We can assume that the underestimation of mass is negligible.

By using our model, we can derive the ionising photon flux as %
\begin{equation}
  \label{equ:fuv_grid}
  {\rm F_{UV}} = (\Delta x)^3\,\beta_{2} \sum_i n_e^{2}(x_i,y_i,z_i),
\end{equation}
with $\beta_{2}$ the recombination coefficient of the second energy
level of H. It is still not known if the nebula is density or
ionisation bounded, therefore we have to keep in mind that this could
be a lower limit. We find $\rm Log(F_{UV})=44.5$,
which corresponds to a supergiant of spectral type later than B3. This
is too cold compared with S61's star \citep{1997Crowther,
  1999Pasquali}. Note that the recombination time for such a nebula
would be typically of some thousands years, implying that LBV
variability from the ionising source would be negligible. This may
indicate that the nebula is density-bounded and that part of the
stellar UV flux escapes from the nebula.
\begin{figure}
  \includegraphics[width=1\columnwidth]{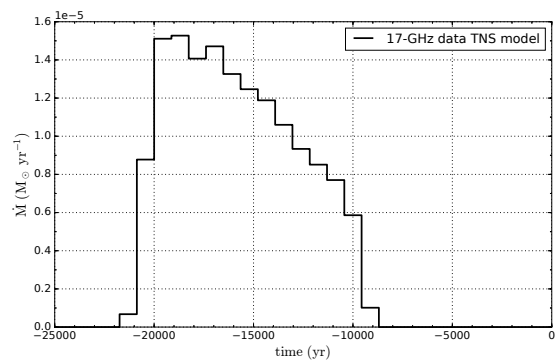}
  \caption{Mass-loss rate of S61 as derived from fitting the circumstellar nebula. Each bin of the histogram corresponds to $\sim850\,\rm yr$. The plot does not contain information about the present-day mass-loss rate.}
  \label{fig:s61massloss}
\end{figure}

\section{Discussion}
\label{sec:discuss}

\subsection{The mass-loss history}

Starting from our best model in Section \ref{sec:model}, we now derive the mass-loss
history of S61 with high temporal resolution, keeping in mind that we could not constrain the electron density distribution. 

If we know the expansion velocity of this nebula, each voxel of our
datacube corresponds to a kinematical age. For instance, in the case
of S61 the expansion velocity is 27 $\rm km\,s^{-1}$ \citep{Weis2003}
and each voxel in the model of the 17 GHz data has a 1-D size equal to
0.1 arcsec, which, at the assumed distance of the LMC (48.5 kpc)
corresponds to 7.3$\times10^{11}\,\rm km$ and therefore to a
kinematical interval of $\sim850\,\rm yr$. We know the mass in each
voxel and therefore we can derive the average mass-loss rate $dM/dt$
in intervals of $850\,\rm yr$. According to the shell geometry adopted
for the S61 nebula, we can assume that the star has lost mass
isotropically. We can finally integrate $dM/dt$ over shells of radius
$r$ and thickness $dr$, where $r$ can vary between 0 and
$\frac{N}{2}+1$ ($N$ is the number of pixels in each dimension of the
cube). The resulting mass-loss rates for S61 are shown in
Fig.~\ref{fig:s61massloss}. In this particular example, the peak of
the mass-loss has occurred at epoch $\sim 19100$ yr with a rate of
$1.5\times10^{-5}\,\rm \Msun\,yr^{-1}$.  However, we note that the finite resolution due to the synthesised beam may mean that the real distribution is less smooth.

The mass-loss rates derived in Fig. \ref{fig:s61massloss} are
consistent with our non-detection of the stellar wind in the radio
maps.  If we assume a spherical mass-loss for the star and then the
model in \citet{1975PF}, with a terminal velocity of 250
$\rm km\,s^{-1}$ \citep{1997Crowther}, an electron temperature of 6120
K \citep{1999Pasquali} and a flux density equal to 3 times the noise
in the map at \hbox{17\,GHz}, we derive a 3$\sigma$ upper limit of
$\sim 2.3\times10^{-5}$ $\rm \Msun\,yr^{-1}$, for a fully ionised wind
with solar abundances. This upper limit is consistent with the
mass-loss rate of $\sim 1.1\times10^{-5}$ $\rm \Msun\,yr^{-1}$ derived
by \citet{1997Crowther}, which would be within the distribution in
Fig. \ref{fig:s61massloss}. The value by \citet{1997Pasquali} of
$\sim 2.2\times10^{-5}$ $\rm \Msun\,yr^{-1}$ derived from H emission
lines seems inconsistent with the radio
non-detection. 

We now compare the mass-loss history with the empirical mass-loss rate, as predicted for O, B normal supergiants following the procedure described in \citet{2000Vink}.
We first assume the stellar parameters by \citet{1997Crowther} ($T_{eff}=27.6\,\rm kK$, $\log\left(\frac{L}{L_{\odot}}\right)=5.76$ and $v_{\infty}=250\,\rm km\,s^{-1}$), a metallicity of $Z=0.5\,\rm Z_{\odot}$ and an initial 
stellar mass of $\sim60\,\rm M_{\odot}$ \citep[according to the evolutionary tracks by][]{1992Schaller}. The small nebular mass derived in the previous section suggests that the star has a stellar mass similar to its initial value and then the mass-loss rate is comparable to the one relative to the O, B main sequence stars \citep[the reduced stellar mass of LBVs causes a strong increase in the mass-loss rate with respect to normal O, B supergiants, as showed by][]{2002Vink}. The empirical mass-loss rate derived with the mentioned parameters is $6.6\times10^{-6}\,\rm M_{\sun}\,yr^{-1}$, which is close to the average value of the distribution in Fig. \ref{fig:s61massloss}. The consequence of this result is either that the mass-loss occurred with a constant wind (constant density model) or the mass-loss rates varied due to excursion through the bi-stability jump (power-law electron density model). In this latter case the bi-stability jump for S61 would occur at $T_{eff}\sim23.5\,\rm kK$, with a mass-loss rate of $3.3\times10^{-5}\,\rm M_{\odot}\,yr^{-1}$, which is consistent with the peak in Fig. \ref{fig:s61massloss} within a factor of $\sim2$. In both cases we can probably exclude that S61 lost mass through eruptions, as normal stellar winds perfectly explain the observations. If we instead use the stellar parameters by \citet{1997Pasquali} ($T_{eff}=36.1\,\rm kK$, $\log\left(\frac{L}{L_{\odot}}\right)=6.1$), the derived empirical mass-loss rate is $7.0\times10^{-5}\,\rm M_{\sun}\,yr^{-1}$, which is far higher than our observational values derived in the previous section.

\subsection{Extinction map and nebular dust}

We have derived the extinction map of S61 by comparing pixel-by-pixel the highest-resolution radio image (\hbox{17\,GHz}) with the {\it HST\/} $\rm H \rm \alpha$ image, as the two emissions trace the same gas (Paper I). According to \citet{1984Pottasch}, if the optical $\rm H \rm \alpha$ emission is due to the de-excitation of the recombined H atom and the radio continuum emission to free-free encounters, one can determine the extinction of the optical line by comparing the two brightnesses, as

 \begin{equation}
 F_{\nu (expected)} = 2.51\times10^{7}T_{e}^{0.53}\nu^{-0.1}YF_{\rm H_{\beta}} \hspace{0.5cm} \rm [Jy]
\label{predicted}
\end{equation}
where $T_{e}$ is the electron temperature of the nebula in units of K,
$\nu$ is the radio frequency in GHz, Y is a factor incorporating the
ionised He/H ratios (assumed to be 1, as in Paper I) and
  \begin{equation}
 F_{\rm H_{\beta}}=\frac{1}{2.859}\left(\frac{T_{e}}{10^{4}}\right)^{0.07}\,F_{\rm H_{\alpha}}
 \end{equation}
for the theoretical Balmer decrement. 

\begin{figure}
  \begin{minipage}{1\linewidth}
    \centering
    \includegraphics[width=\textwidth]{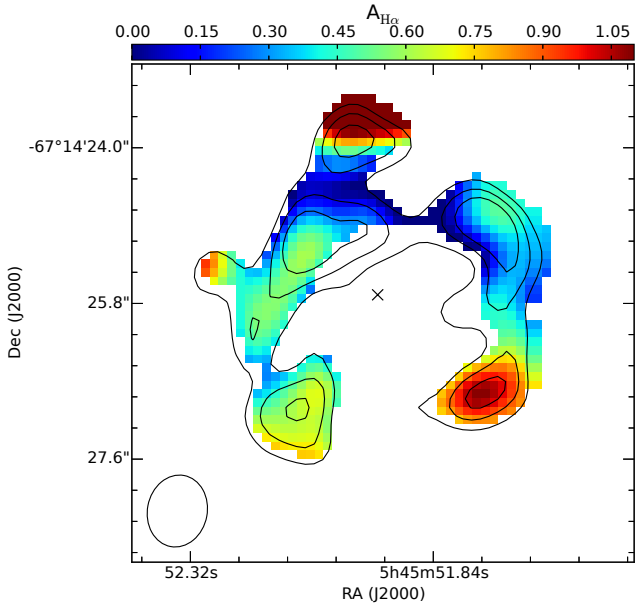}
  \end{minipage}
  \caption{Extinction map in $ \rm H \rm \alpha$, as derived by
    comparing the $ \rm H \rm \alpha$ recombination line and the
    centimetre (\hbox{17\,GHz}) emission above $5\sigma$. The central
    star was masked with a circular aperture in the optical image. The
    black contours are 5, 7 and 9$\sigma$ levels of the radio emission
    and the black ellipse is the resolution. The black cross
    represents the position of the star, according to the Simbad
    Database. }
  \label{fig:ext1}
\end{figure}

We re-gridded the {\it HST\/} image to the same grid of the radio map
and converted it to Jy pixel$^{-1}$
unit. 
We convolved the optical image with the radio beam (elliptical
Gaussian with HPBW as in Table~\ref{tab:ATCA2maps}). Adopting as
electron temperature $6120\,\rm K$ \citep{1999Pasquali}, we derive the
expected radio map from the $\rm H \rm \alpha$ recombination-line
emission. Keeping in mind that we want to estimate the expected
free-free emission from the optical line in the nebula, we masked the
$ \rm H \rm \alpha$ emission from the star. Finally, the extinction
map in $ \rm H \rm \alpha$ was derived as
$2.5\,\log(F_{\rm 17\,GHz\,(obs)}/F_{\rm 17\,GHz\,(expected)})$ in
every common pixel with brightness above 5$\sigma$, where $\sigma$ was
computed by summing in quadrature the noise in the maps and
calibration uncertainties (however negligible). As a result of this
procedure, we obtained the extinction map illustrated in
Fig.~\ref{fig:ext1}.
\begin{figure}
 
 \begin{minipage}{1\linewidth}
\centering
  \includegraphics[width=\textwidth]{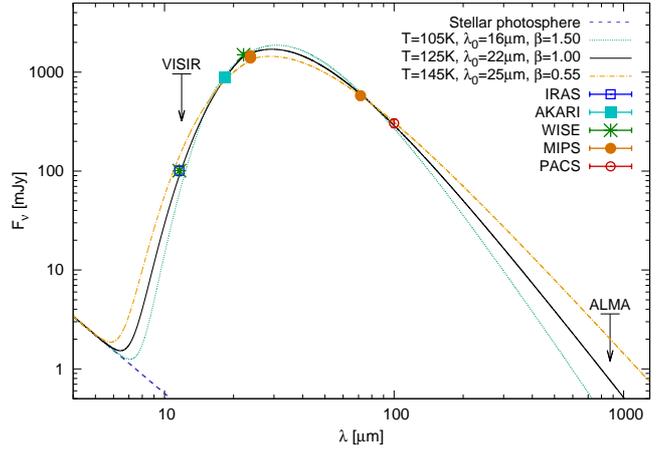}
\end{minipage}
\caption{Flux density distribution of S61 derived by using the flux densities in the IR catalogues. The continuum lines represent the fit grey-body functions obtained for three values of the parameter $\beta$. The 3$\sigma$ upper limit spatially integrated flux densities at the VISIR and ALMA observing wavelengths (\hbox{11.88\,$\rm \mu m$} and \hbox{875\,$\rm \mu m$}, respectively) are also shown.}
  \label{fig:sed}
\end{figure} 
Small extinction due to dust is evident across the whole region. The range of values for A$_{\rm H\alpha}$ across the nebula is between $\sim$0.1 and 1.09. The maximum value for A$_{\rm H\alpha} $ is 1.8 and corresponds to the spur-object in the Northern (upper) part.  

To derive a range of possible characteristic temperatures for the dust that extinguishes the optical emission, we fitted the flux density distribution from the mid- to the far-IR. We consulted the IR catalogues with the VizieR tool \citep{2000Ochsenbein} at the position of S61, and we extracted the flux densities in the \emph{WISE} bands W3 and W4 \citep{2012aCutri,2012bCutri}, \emph{AKARI} L18W at 18 $\rm \mu m$ \citep{2010aIshihara, 2010bIshihara}, \emph{Spitzer} MIPS at 24 and 70 $\rm \mu m$ \citep{2010aArdila,2010bArdila} and \emph{Herschel} PACS at 100 $\rm \mu m$ \citep{2013Meixner}. We fitted a single-temperature greybody with power-law opacity index $\beta$ at longer wavelengths and constant opacity at shorter wavelengths \citep[e.g.][]{1993backman},
\begin{equation}
F_\nu(\lambda)\propto\tau(\lambda)B_{\nu}(\lambda),\quad
\tau(\lambda) = \left\{\begin{matrix}
\tau_0 & \lambda \leq \lambda_0\\
\tau_0\left({\lambda\over\lambda_0}\right)^{-\beta} & \lambda > \lambda_0
\end{matrix}\right.
\end{equation}
We found a range of characteristic temperatures between $\sim\!105$ to
$145\,\rm K$ by varying the parameter $\beta$ (we explored the cases
for $\beta = 0.55, 1$ and $1.5$) and $\lambda_{0}$ (between 18 and 25
$\rm \mu m$). The modified black-body that best fits the data is the
one represented with a dark continuum line in the figure, with
$\beta=1.0$. Note that $\beta = 1.0$ implies either the existence of
relatively large grains or different dust components (of different
temperatures). The latter is usually observed in Galactic LBVNe and
the temperature of the dust decreases with increasing distance from
the star \citep[e.g.][]{1997Hut, 2016Buemi}.

According to the flux density distributions in Fig.~\ref{fig:sed}, the
expected flux densities $F_{\nu}$ at the VISIR PAH2$\_$2 and Q1
central wavelengths ($11.88$ and $17.65\,\rm \mu m$) are:
$105\pm40\,\rm mJy$ and $764\pm50\,\rm mJy$, respectively. If this
emission arises from a point-like source close to the star as observed
in other candidate LBVs \citep[e.g. G79.29+0.46,][]{2014Agliozzo}, we
would have detected it with VISIR. We deduce then that the dust is
spread out over the nebula, at angular scales that our observations
were not sensitive to. Similarly, we have derived the expected ALMA
flux density at \hbox{343\,GHz}: $0.8^{+1.2}_{-0.5}\,\rm mJy$. Even
for the most favourable case for the dust ($\beta \sim0.5$, corresponding
to optically thick large grains), the sensitivity achieved with only
one execution was not sufficient to detect the thermal emission at
sub-mm wavelengths. In fact, with an expected flux density of
$2\,\rm mJy$ (case $\beta=0.55$), spread across 16.7 ALMA synthesised
beams, the average brightness would be $0.12\,\rm mJy\,beam^{-1}$ and
a sensitivity of $40\,\rm \mu Jy\,beam^{-1}$ was needed for a
$3\sigma$ detection. Note that the extrapolated flux density at the
ALMA frequency is consistent with the upper limit derived from the
map: the rms-noise ($72\,\rm \mu Jy$), integrated over the area
corresponding the ionised nebula, yields a 3$\sigma$ upper limit of
$\sim 3.6\,\rm mJy$. Using the flux density extracted from the best
fit (case $\beta=1.0$) at the ALMA frequency \hbox{343\,GHz} (see
Fig.~\ref{fig:sed}), we derived a dust mass $M_{d} = 5\times10^{-3}$
to $2.7\times10^{-2}\,\rm \Msun$, considering that
$M_{d}=S_{\nu}\,D^{2}/(B_{\nu}(T)\,\kappa_{\nu})$, and assuming
$\kappa_{343\rm \,GHz}= 1\rm \,cm^{2}\,g^{-1}$. This means a low
gas-to-dust ratio for the LMC. It suggests that the sub-mm emission
might be even lower than that computed from the flux density
distributions based on the mid- and far-IR data.
  
The extinction map resembles the dusty nebula around the Galactic LBV
IRAS 18576$+$034 \citep{2010Buemi}. This was also observed with VISIR
in the filters PAH2$\_$2 and Q1. They derived for the mid-IR nebula a
dust component of temperature ranging from 130 to 160 K. IRAS
18576$+$034 has a mid-IR nebula of 7 arcsec diameter, corresponding to
0.35 pc at the distance of 10 kpc. It has a physical size which is
about half that of S61, but in the sky the two sources have similar
angular size. We rescaled the IRAS 18576$+$034 VISIR maps to the
distance of the LMC, and we derived from the maps a mean value of
$0.3\,\rm mJy\,pixel^{-1}$ in the PAH2$\_$2 filter and
$1.3\,\rm mJy\,pixel^{-1}$ in the Q1 filter. This means that with our
VISIR observations, we would have detected at $\sim3.5 \sigma$ in the
PAH2$\_$2 filter image a nebula like IRAS 18576$+$034. The sensitivity
reached in band Q1 would have not been sufficient to detect the
nebula. \citet{2010Buemi} derived for IRAS 18576$+$034 a dust mass of
$\sim 4.5-6.5\times10^{-3}\,\rm \Msun$ (depending on the assumed dust
composition) and \citet{2005Umana} derived a mass of
$\sim 2\,\rm \Msun$ for the ionised gas. This suggests the dust
content in the nebula around S61 is similar in mass to that estimated
in IRAS 18576$+$034. Conversely, the ionised mass in S61 is only a
small fraction (1/20th) of the mass in the IRAS 18576$+$034 nebula,
despite S61's nebula diameter ($\sim 1.2\rm \,pc$) being about 3.5
times bigger than IRAS 18576$+$034 ($\sim 0.35\rm \,pc$). We recall, however, that IRAS 18576$+$034 has an estimated bolometric luminosity higher than S61 \citep[$\log\left(\frac{L}{L_{\odot}}\right)=6.4$,][]{2001Ueta} and the mass, through the mass-loss' quadratic dependence on luminosity, has a stronger effect than the metallicity. The inner
shell around Wray 15-751, which has a luminosity similar to S61, extends up to $1\rm \,pc$, similar to S61 and
has gas and dust masses of $1.7\pm0.6\,\rm \Msun$ and
$\sim3.5\times10^{-2}\,\rm \Msun$, respectively \citep{2013Vam}, more
massive than the S61 nebula. This may suggest that S61's mass-loss 
has been less efficient over time than the mentioned Galactic
LBVs. The dust production does not seem significantly
different. However, a potential issue for this comparison is the larger uncertainty of the Galactic LBVs distances than those of the Magellanic objects.

\section{Conclusions}
\label{sec:summary}
In this work we presented high spatial resolution observations from
the radio to the mid-IR of the nebula associated with the candidate
LBV S61. It was detected only in the centimetre band. The nebula has a
morphology resembling a shell, as in the optical, but in the radio
there is more sub-structure. The emission mechanism is optically thin
free-free, as evidenced by the spectral index map, although there are
regions that suggest self-absorption.

We developed and made publicly available a code in Python that permits
to model the 3-D electron density distribution and to derive the mass
in the nebula. We tried different geometries for the shell (truncated
Gaussian, constant density and power-law $n_e\propto r^{-1}$ and
$\propto r^{-2}$) and we found that at least three of these geometries
give similar-quality results. For all the well-fitting models, the
derived ionised mass is always about 0.1 M$_{\odot}$, which is an order of magnitude smaller than previous estimates and also a few factors smaller 
than the mass of similar Galactic objects. The
nebula is very likely density bounded, meaning that part of the
stellar UV flux escapes from the nebula. As an application of our
modelled electron density distribution, we also show how to derive the
mass-loss history with high-temporal resolution ($\sim$ 850 yr). The
derived mass-loss rates are consistent with the empirical mass-loss rate for S61, implying that the nebula was likely formed by stellar winds, rather than eruptive phenomena. The present-day mass-loss is $\lesssim2\times10^{-5}$
$\rm \Msun\,yr^{-1}$.

Based on the extinction map derived from the radio map and the
$\rm H_{\alpha}$ {\it HST\/} image, we have explored the possibility
that the nebular regions with higher spectral index are dusty, by
means of high-resolution mid-IR and sub-mm observations. We did not
detect any point-like source, or compact regions associated with the
clumps, neither with VISIR nor with ALMA. The fit of the IR flux
distribution from space telescope observations suggest the
presence of dust with a range of characteristic temperatures of
$125\pm 20\,\rm K$ and dust mass $M_{d}$ of $10^{-3}$ to
$10^{-2}\,\rm \Msun$. Based on the observations with VISIR and ALMA,
we exclude that the IR emission arises from a point-like source.  The
dust producing the infrared emission observed by space telescopes
must be searched for within the angular scales of the ionised gas
($\sim$1--5 arcsec). The dust is distributed in
an optically thin configuration over the radio nebula, but not
uniformly, as shown in the extinction map.  The VISIR observations did
not reach the required sensitivity to detect such extended thermal
emission. With the ALMA observations, we obtained better sensitivity
to study thermal emission, but still the nebula was not detected. We
estimate that the thermal emission could be detected by deeper ALMA
observations in the future, including 7m antennas to enhance
sensitivity on larger angular scales.

\section{Supplements}
\label{sec:supplements}
\rhocube\ \citep{rhocube} is a general-use, stand-alone code and is
distributed as such in the following git repository:
\url{https://github.com/rnikutta/rhocube} . In the spirit of
scientific reproducibility we also share with the reader all scripts
and supplementary codes that we have used in this work, specifically
the MCMC sampling and Bayesian inference functions that make use of
\rhocube, the functions to compute the ionised mass within a density
model and the mass-loss history, and some plotting routines. A git
repository for this manuscript, holding all supplementary files
including the data FITS files, is accessible at:
\url{https://github.com/rnikutta/s61-supplements}

\section*{Acknowledgements}
The authors wish to thank the referee for their valuable suggestions that helped to improve the presentation of this work.  
We acknowledge support from FONDECYT grant No. 3150463 (CA) and
FONDECYT grant No. 3140436 (RN), and from the Ministry of Economy,
Development, and Tourism's Millennium Science Initiative through grant
IC120009, awarded to The Millennium Institute of Astrophysics, MAS (CA
and GP). We wish to thank the ESO Operations Support Center for the support received for the VISIR data reduction.

This paper makes use of the following ALMA data:
ADS/JAO.ALMA$\#$2013.1.00450.S. ALMA is a partnership of ESO (representing
   its member states), NSF (USA) and NINS (Japan), together with NRC
   (Canada) and NSC and ASIAA (Taiwan) and KASI (Republic of Korea), in
   cooperation with the Republic of Chile. The Joint ALMA Observatory is
   operated by ESO, AUI/NRAO and NAOJ. 

In addition, this research: is based on observations made with ESO telescopes at the La Silla Paranal Observatory under programme ID 095.D-0433; makes use of data products from the Wide-field Infrared Survey Explorer, which is a joint project of the University of California, Los Angeles, and the Jet Propulsion Laboratory/California Institute of Technology, funded by the National Aeronautics and Space Administration; made use of the VizieR catalogue access tool, CDS, Strasbourg, France; made use of the NASA/ IPAC Infrared Science Archive (IRSA), which is operated by the Jet Propulsion Laboratory, California Institute of Technology, under contract with the National Aeronautics and Space Administration; made use of the SIMBAD database,
operated at CDS, Strasbourg, France; made use of Montage, funded by the National Aeronautics and Space Administration's Earth Science Technology Office, Computation Technologies Project, under Cooperative Agreement Number NCC5-626 between NASA and the California Institute of Technology. Montage is maintained by IRSA.

\bsp

\appendix

\section{Brief introduction to RHOCUBE}
\label{sec:usage-rhocube}

\subsection{Principles}

\rhocube\ computes a 3D density field \rhoxyz\ on a Cartesian,
right-handed grid, with $x$ pointing to the right, $y$ pointing up,
and $z$ pointing to the viewer. The grid resolution is set by the user
upon model instantiation. Several models with density distributions of
common interest are provided, and new ones can be easily added by the
user. At the time of writing, the (mnemonically named) provided models
are: \emph{PowerLawShell}, \emph{TruncatedNormalShell},
\emph{ConstantDensityTorus}, \emph{ConstantDensityDualCone},
\emph{Helix3D}. The latter takes as the \emph{envelope} parameter the
imaginary surface on which the helical tube spirals (\emph{dual cone}
or \emph{cylinder}). The models are implemented as Python classes, and
all inherit basic functionality, such as e.g. 3D rotations and $x,y$
offsets, from a common class \emph{Cube}.

\begin{figure*}
  \centering
  \includegraphics[width=\textwidth]{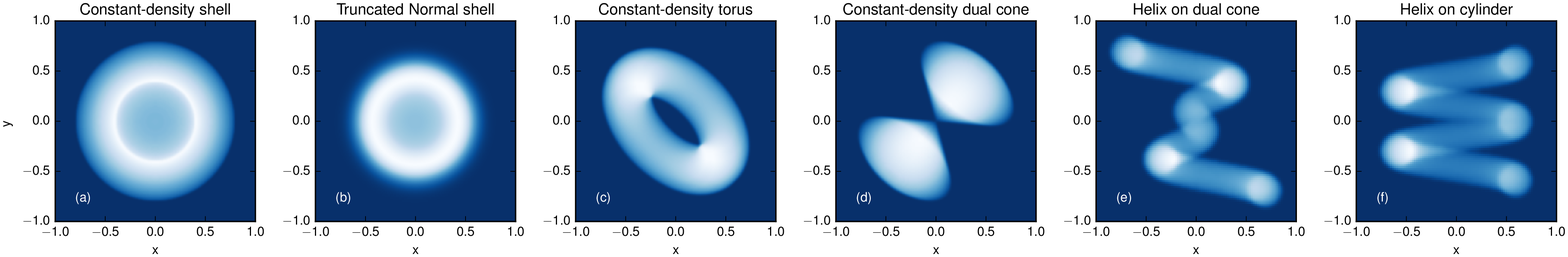}
  \caption{Example gallery of some $z$-integrated model density maps
    that can be generated with \rhocube. $x$ and $y$ are in relative
    linear units. The colour map is scaled to the image maxima. (a)
    Constant-density shell with four free parameters: inner and outer
    shell radii, and the centre offsets \xoff\ and \yoff. (b) A
    truncated Gaussian-density shell with 6 free parameters: radius,
    width of the Gaussian, left and right cut-off radii, two
    offsets. (c) A constant-density torus with six free parameters:
    radius, cross-section, two tilt angles, two offsets. (d) A
    dual-cone with six free parameters: height, opening angle, two
    tilts, two offsets. (e) Helical tube on the surface of a dual
    cone, with the free parameters radius at top of the cone, number
    of turns taken from origin, tube diameter, three tilts, two
    offsets. (f) Like (e) but on the surface of a cylinder. An
    arbitrary combination of models, with relative total masses as
    free parameters, is also possible.}
  \label{fig:rhocube_gallery}
\end{figure*}

Every model has a number of free parameters, e.g. for
\hbox{\emph{PowerLawShell}} the inner and outer radii \texttt{rin} and
\texttt{rout}, and the radial power-law index \texttt{pow}. Two
lateral offsets \xoff\ and \yoff\ to de-center the density
distribution in the image plane are available to all the models, as
are the rotation angles \texttt{tiltx}, \texttt{tilty},
\texttt{tiltz}, which, when provided, rotate the entire 3D density
field about the respective axes. They are of course ineffectual for
spherically symmetric density distributions. Figure
\ref{fig:rhocube_gallery} shows a few examples of integrated maps that
can be computed with the code.

\subsection{Usage}
\label{sec:usage}

The workflow with \rhocube\ is straight-forward:
\begin{itemize}
\item Instantiate a model (e.g. \texttt{PowerLawShell})
\item Call the instance with a set of parameter values
\item Retrieve/access the 3D density cube and/or 2D $z$-integrated image
\end{itemize}
Instantiating a model generates the Cartesian grid (with requested
resolution), and provides it with general methods to manipulate the
density distribution (e.g. 3D rotations and shifts in \xyz). Once the
model is created, it can be called any number of times with a set of
numerical arguments which are the values that the free model
parameters should assume. Each call computes the corresponding 3D
density field and also integrates that field along the $z$ axis,
storing the resulting 2D image as a member of the model instance. If
the \texttt{smoothing} parameter was set to a float value, then
\rhoxyz\ will be smoothed with a 3D Gaussian kernel (see
Sec. \ref{sec:smoothing}). Listing \ref{'lst:inst'} show a
simple example instantiating and calling a simple model.
\lstset{
name=rhocubeusage,
language=Python,                             
basicstyle=\scriptsize\ttfamily,                   
frame=false,                             
captionpos=t,                           
breaklines=false,                        
showspaces=false,                       
}
\begin{lstlisting}[name=inst,caption={Instantiating and calling a model.},label='lst:inst']
from models import PowerLawShell as PLS
mod = PLS(101,exponent=0.,smoothing=1.) # 0. for const.dens. shell
args = (0.1,0.7,0,0,None)  # rin,rout,xoff,yoff,weight
mod(*args)
\end{lstlisting}
The resulting \rhoxyz\ is now in \texttt{mod.rho} (as 3D array), and
the summation along $z$ direction is in \texttt{mod.image}. If you
wanted to vary the parameters of this model:
\begin{lstlisting}[name=inst,label='lst:inst']
args = (0.3,0.5,0,0.3,None)  # rin,rout,xoff,yoff,weight
mod(*args)
\end{lstlisting}

If a \texttt{transform} function is provided during instantiation, it
will be applied to \rhoxyz\ before integration (see
Sec. \ref{sec:transform}). If during calling the \texttt{weight}
parameter is set to a float value, the sum \hbox{$\sum_i \rhoxyz$} (of
the possibly transformed \rhoxyz) over all voxels will be normalized to
that value. Listing \ref{'lst:call'} shows an example.
\begin{lstlisting}[name=call,caption={Using \texttt{transform}
function and \texttt{weight} parameter.},label='lst:call']
from rhocube import PowerTransform
mod = PLS(101,exponent=-1.,smoothing=1.,transform=PowerTransform(2.))
args = (0.1,0.7,0,0,1.)  # rin,rout,xoff,yoff,weight
mod(*args)
print model.rho.sum()
  1.0   # the specified 'weight'
\end{lstlisting}

\subsection{Special methods}
\label{sec:special}

\subsubsection{The 'transform' function}
\label{sec:transform}
An optional transform function $f(\Varrho)$ can be passed as an
argument when creating a model instance. The transforms are
implemented as simple Python classes.
A transform will be applied to the density field before
$z$-integration, i.e. \hbox{$\int\! \dif z\, f(\rhoxyz)$} will be
computed.
In our case for instance, squaring of the electron number density
before integration is required by Eq.~\eqref{equ:emmeasure}.
If the supplied $f(\Varrho)$ class also provides an inverse function,
e.g. \hbox{$f^{-1} = \sqrt{\cdot}$ when $f=(\cdot)^2$}, then the
entire 3D cube \hbox{\rhoxyz} with correct scaling can also be
computed and accessed by the user.
Some common transform classes are provided with \rhocube,
e.g. \texttt{PowerTransform}, which we use for the squaring mentioned
above (with argument \texttt{pow=2}), or \texttt{LogTransform} which
computes a \texttt{base}-base logarithm of \rhoxyz. Another provided
transform is \texttt{GenericTransform} which can take any
parameter-free \textsc{Numpy} function and inverse (the defaults are
\texttt{func='sin'} and \texttt{inversefunc='arcsin'}). Custom
transform functions can be easily added.

\subsubsection{Smoothing of the 3D density field}
\label{sec:smoothing}

Upon instantiating a model, the \texttt{smoothing} parameter can be
specified. If \texttt{smoothing} is a float value, it is the width (in
standard deviations) of a 3D Gaussian kernel that \rhoxyz\ will be
convolved with, resulting in a smoothed 3D density
distribution. Smoothing does preserve the total \hbox{$\sum_i \rhoxyz$},
where $i$ runs over all voxels. \texttt{smoothing=1.0} is the default,
and does not alter the resulting structure significantly. If
\texttt{smoothing=None}, no smoothing will be applied.

\subsection{Providing own density distributions}
\label{sec:own-density}

The \texttt{Cube} class provides two convenience objects and methods
to compute the 3D density \rhoxyz, which can (but don't need to) be
utilised by the actual model upon instantiation. The two methods are
\texttt{computeR} and \texttt{buildkdtree}.

\subsubsection{X,Y,Z coordinate arrays}
By default, 3D Cartesian coordinate grids \texttt{X},\texttt{Y},and
\texttt{Z} are computed upon instantiation of the \texttt{Cube} class,
and each holds the $x$ or $y$ or $z$ coordinates of the voxel centers,
in fractional units of a cube with extent [-1,1] along every
axis. They can be used to compute arbitrary dependencies \rhoxyz.

\subsubsection{Distance array}
If \texttt{computeR=True} is passed to \texttt{Cube} during model
instantiation, then the class will also compute a 3D radius grid
\texttt{R(x,y,z)}, i.e. a cube of $\rm npix^3$ voxels, each holding
its own radial distance from the cube center. This $R$ cube can then
be used inside the model to compute a distance-dependent density as
$\Varrho\!(R)$. This method is used in all azimuthally-symmetric
models that come with \rhocube, e.g. \texttt{PowerLawShell} and
\texttt{TruncatedNormalShell}.

Below we show in a simple example how one can construct a custom 3D
density model that computes a spherically symmetric $\Varrho\!(R)$
which varies as the cosine of distance, i.e.
\hbox{$\Varrho\!(R) \propto \cos(R)$.}
\begin{lstlisting}[name=cosineshell,caption={},label='lst:cosineshell]
class CosineShell(Cube):
    def __init__(self,npix):
        Cube.__init__(self,npix,computeR=True)
    def __call__(self):
        self.get_rho()
    def get_rho(self):
        self.rho = N.cos(self.R)
        self.apply_rho_ops() # shift, rotate3d, smoothing
\end{lstlisting}
You can then use it simply like this:
\begin{lstlisting}[name=cosineshelluse,caption={},label='lst:cosineshelluse]
import CosineShell
mod = CosineShell(101)  # 101 pixels along each cube axis
mod()  # this model has no free parameters
\end{lstlisting}
Please see the built-in model classes for more details and ideas.

\subsubsection{k-d tree}
\rhocube\ also supports non-symmetric or irregular density
distributions. One example might me the (also provided) model for a
helix that winds along some prescribed parametric curve. For fast
computation of all voxels within some orthogonal distance from the
parametric curve (i.e. within a 'tube'), we utilise the second helper
method in \texttt{Cube}, namely a k-d tree \citep{1975bentley}. The
Helix3D model works like this:
\begin{lstlisting}[name=helix,caption={},label='lst:helix]
from models import Helix3D
mod = Helix3D(101,envelope='dualcone')  # 'cylinder' also available
args = (0.8,1,0.2,0,0,90,0,0,None) # h,nturns,tilt_x/y/z,
                                   # xoff,yoff,weight
mod(*args)
\end{lstlisting}
Note that the initial building of the k-d tree is a
$\mathcal{O}(n \log^2{}n)$ operation. The subsequent lookups are then
much faster. Please see the \texttt{Helix3D} class for more details of
the implementation.

\section{Bayesian parameter inference}
\label{sec:bayparinf}

\subsection{Conditional probability and Bayes' Theorem}
Estimating the most likely physical parameters (inputs) of a model
whose output is compared to observed data, is by far the most common
scenario of Bayesian statistics.
In our case, the inputs are the geometrical parameters of the modelled
3D electron density distribution $n_e(x,y,z)$, and the outputs are the
modelled \emph{emission measure} (EM) maps.
They are compared to the observed EM maps of a given LBV shell.

We must vary the free model parameters, with the objective of
minimizing the deviation of the resulting model EM map and the data EM
map (for instance $\chi^2$ if the data errors can be assumed
Gaussian).
We desire not only to estimate the best-fit parameters, but to
quantify their uncertainty, or the \emph{confidence} that we can have
in the results.
The most natural approach to this common parameter estimation problem
is Bayesian inference.
Using notation borrowed from statistical literature, \emph{Bayes'
  Theorem}
\begin{equation}
  \label{eq:bayes}
  Posterior \equiv p(\vec\theta|\vec D) = \frac{p(\vec\theta)\, p(\vec D|\vec\theta)}{p(\vec D)} \equiv \frac{Prior \times Likelihood}{Evidence}
\end{equation}
provides a straight-forward prescription how to compute the joint
\emph{posterior} probability distribution (PDF) $p(\vec\theta|\vec D)$
of a possibly multi-variate vector of model parameters
$\vec\theta = (\theta_1,\theta_2,\ldots)$, given the observed data
vector $\vec D$.
The posterior PDF distribution is simply a product of a \emph{prior}
PDF $p(\vec\theta)$ (i.e. any knowledge we may have of the model parameter
distribution \emph{before} introducing the data) with the
\emph{likelihood} that the given parameter values generate a model
that is compatible with the data.
For normally distributed errors the likelihood is
$p(\vec D|\vec \theta) \propto \exp(-\chi^2/2)$ \citep[see e.g.][]{Trotta_2008}.

The \emph{evidence} $p(\vec D)$ in Eq. \eqref{eq:bayes} is the
normalisation (integral of the multi-dimensional posterior PDF),
ensuring that the total probability be unity.
For the sole purpose of parameter inference, it is not necessary to
compute the evidence explicitly, since it does not change the
\emph{shape} of the posterior PDF.
It is instead sufficient to re-normalize the posterior to unit volume
a posteriori.
Thus, for parameter inference, only the relation
$p(\vec \theta|\vec D) \propto p(\vec \theta)\, p(\vec D|\vec \theta)$ is relevant.
Of particular interest for the interpretation of results are the
\emph{marginalized} 1D posterior distributions, each integrated over
all model parameters but the one in question
\begin{equation}
  \label{eq:marginalized}
  p(\theta_i|\vec D) = \int\! \dif\theta_1 \dif\theta_1 \ldots \dif\theta_{i-1} \dif\theta_{i+1} \ldots \dif\theta_N\ p(\vec D|\vec \theta).
\end{equation}
Every $\theta_i \in \vec\theta$ is one of the free model parameters,
and every $\theta_j \ne \theta_i$ is a so-called \emph{nuisance
  parameter} when computing the marginalized posterior PDF of
$\theta_i$. Hence the common expression ``marginalize over the
nuisance parameters''. In our application, the marginalized posteriors
are shown in panels (2)--(5) in Fig.~\ref{fig:s61_posteriors}. Panel
(6) shows the posterior PDF of a derived quantity, which in the
Bayesian approach is trivial to compute, and in the ``classical''
approach, impossible.

\subsection{MCMC sampling}
The $N$ model parameters span an $N$-dimensional volume which grows
exponentially with the number of parameters.
It very quickly becomes impractical to sample the entire
volume.
Fortunately, for many problems only small sub-volumes are relevant,
i.e. the likelihood is only high in small regions of the parameter
volume.
Several methods to sample preferentially these highly significant
regions have been proposed.
The best-known is probably Markov-Chain Monte Carlo sampling (MCMC).
A particularly straight-forward MCMC formalism is the
\emph{Metropolis-Hastings algorithm} introduced by
\citet{Metropolis+1953} and later generalized by
\citet{Hastings_1970}.
It can be shown that the proposal joint PDF from which the algorithm
samples eventually converges towards the sought-after target posterior
PDF.
MCMC thus generates a chain of samples (for every model parameter
$\theta_i$), whose histogram \emph{is} the marginalized posterior
$p(\theta_i|\vec D)$.

A common way to characterise the marginal PDFs is to compute the
median (0.5 percentile of the cumulative distribution function, CDF),
and as the confidence interval (or ``credible interval'') the
boundaries of an inter-percentile range.
For Gaussian PDFs this can be the $\pm 1\sigma$ interval around the
median, i.e. the range [0.158--0.841] of the CDF.
For (slightly) asymmetric PDFs, an inter-quartile range is often used,
i.e. [0.25--0.75].
While the posteriors in our application are not always Gaussian, for
consistency we will report as the confidence interval the
$\pm 1\sigma$ range around the median throughout.

\label{lastpage}

\end{document}